\documentclass[a4paper,11pt]{article}
\pdfoutput=1 

\usepackage{jheppub} 

\usepackage[T1]{fontenc} 

\title{\boldmath Stochastic Baryogenesis}

\author[a]{Yi-Peng Wu,}
\author[a, b]{and Kalliopi Petraki}

\affiliation[a]{Laboratoire de Physique Th\'{e}orique et Hautes Energies (LPTHE), \\
	UMR 7589 CNRS \& Sorbonne Universit\'{e}, 4 Place Jussieu, F-75252, Paris, France}
\affiliation[b]{Nikhef, 
Science Park 105, 1098 XG Amsterdam, The Netherlands}

\emailAdd{ywu@lpthe.jussieu.fr}
\emailAdd{kpetraki@lpthe.jussieu.fr}

\abstract{
	Using a multi-field stochastic approach, we investigate the vacuum expectation value (VEV) during inflation of a scalar field charged under a mildly broken global $U(1)$ symmetry that can play the role of baryon or lepton number, or possibly a dark baryon number or a combination of the three. Even for a CP invariant Lagrangian, the stochastic distribution of inflationary VEVs in general breaks CP spontaneously, allowing for successful baryogenesis via the Affleck-Dine mechanism. 
	For the Hubble scale during inflation $H_\ast$ as high as $10^{13}$ GeV, we show that the post-inflationary relaxation of the charged scalar with stochastic initial conditions can explain the observed baryon asymmetry, and that a charged scalar mass at the order of $H_\ast$  is favored by the isocurvature constraints.
}


\begin{document} 
\maketitle
\flushbottom

\section{Introduction}
\label{sec:intro}



Inflation creates a nearly perfect de Sitter background spacetime in the standard single-field framework and this prediction followed by the slow-roll paradigm has been strongly supported by measurements of the cosmic microwave background anisotropy \cite{Akrami:2018odb}. The Hubble parameter during inflation, $H_\ast$, serves as the fundamental scale that uniquely characterizes such a de Sitter background. With the current constraint $H_\ast < 6.6 \times 10^{13}$ GeV (for the pivot scale at $0.002\, \mathrm{Mpc}^{-1}$), $H_\ast$ seems to be a natural unit for measuring the size of quantum fluctuations during inflation or the vacuum expectation value (VEV) of quantum fields. For a scalar field with a mass $m \lesssim H_\ast$, a non-trivial VEV inevitably occurs due to the condensation of long-wavelength fluctuations against the high energy background expansion, and this process can be formally described by the stochastic inflation approach \cite{Starobinsky:1994bd,Starobinsky:1986fx,Rey:1986zk,Sasaki:1987gy,Nambu:1988je,Morikawa:1989xz,Linde:1991sk,Mollerach:1990zf}. 
\footnote{Stochastic inflation is in particular a compelling method for tackling over non-perturbative quantum corrections to observables residing in inflationary correlation functions \cite{Vennin:2015hra,Finelli:2008zg,Tsamis:2005hd,Finelli:2010sh,Garbrecht:2013coa,Garbrecht:2014dca,Onemli:2015pma,Fujita:2013cna,Fujita:2014tja,Assadullahi:2016gkk,Vennin:2016wnk,Pinol:2018euk,Pinol:2020cdp,Kitamoto:2018dek}.} 

An extremely large VEV developed during inflation, possibly far above the electroweak scale of the Standard Model of particle physics, has important applications to scenarios of generation of the observed baryon asymmetry in our Universe. For example, the pioneering study by Linde \cite{Linde:1985gh} uses inflationary VEVs in the context of the chaotic inflation as initial conditions for baryogenesis via the Affleck-Dine mechanism \cite{Affleck:1984fy} (for a review, see \cite{Dine:2003ax,Dolgov:1991fr}). 
The baryon asymmetry could also convert from the lepton sector with asymmetric chemical potential in the thermal equilibrium induced by the coherent relaxation of a scalar condensate in the post-inflationary epoch \cite{Kusenko:2014lra,Yang:2015ida,Pearce:2015nga,Wu:2019ohx,Inomata:2018htm,Kusenko:2017kdr,Kusenko:2014uta}. Candidates for the scalar condensate include the Standard Model Higgs \cite{Kusenko:2014lra} or an axion field \cite{Kusenko:2014uta}, as long as the field potential is shallow enough to produce a sizable VEV ($\gtrsim H_\ast$) by the end of inflation.  

One of the most remarkable features of a scalar condensate relaxation from inflationary VEVs is that a net baryon number can be generated in a local region of the Universe from a C/CP invariant theory \cite{Dolgov:1991fr,Hook:2015foa}. More precisely, the C/CP symmetry is preserved only at the Lagrangian level, whereas physical solutions in the cosmological background do not necessarily respect the primary symmetry of the Lagrangian (a similar idea as the spontaneous T violation \cite{Lee:1973iz}). In this work, physical solutions of our interest are VEVs of charged scalar fields obtained by the stochastic inflation approach.
We aim to clarify the role of spontaneous C/CP violation among the random distribution of initial VEVs and its implication to the final baryon number asymmetry. 

Previous investigations \cite{Dolgov:1991fr,Hook:2015foa} have demonstrated baryogenesis through a CP invariant model by considering a charged scalar with a mass $m \ll H_\ast$ and with small self-interactions that violate the baryon number with real coupling constants. The scalar potential may or may not exhibit a flat direction (depending on the coupling constants), and the VEV during inflation can be estimated by the massless or the massive (or the self-interacting) formula, depending on the presence of a flat direction. However, the composition of the inflationary VEVs considered in \cite{Dolgov:1991fr,Hook:2015foa}, essentially based on single-field stochastic approach, are insufficient to obtain information associated with the spontaneous CP violation. As we will show in this work, the final baryon asymmetry is determined by the amount of CP violation residing in the initial VEV, with the probability distribution of the latter depending on the coupling constants.

In order to fully take into account the CP violation generated by the inflationary VEVs, we construct a multi-field stochastic formalism, based on the previous works \cite{Starobinsky:1994bd,Adshead:2020ijf}, for a charged scalar field $\phi$ with a mildly broken $U(1)$ symmetry. This global $U(1)$ can be associated with the baryon or lepton number ($B$ or $L$) for the visible-sector particles, the ordinary $B-L$, or it can be a combination of baryon numbers from both the visible and dark sectors (for a review, see \cite{Petraki:2013wwa}). In Section~\ref{Sec.stochastic_inflation}, we firstly choose the coupling constants such that the stochastic equations for the standard decomposition $\phi \rightarrow \phi_{R} + i\phi_{I}$ are fully decoupled between the two components $\phi_{R}$ and $\phi_{I}$. Solutions of $\phi_{R, I}$ are thus given by the single-field formalism \cite{Starobinsky:1994bd}, and these solutions are used to confirm the results of the multi-field formalism in the polar representation $\phi \rightarrow Re^{i\theta}$.

We then extend the study to cases with general coupling constants, where inflationary VEVs can only be obtained by virtue of the multi-field formalism. We provide in Section~\ref{Sec.baryogenesis} both the numerical and the analytical results of the final baryon asymmetry $Y_B$ estimated at the end of reheating. We clarify the difference between the local value $Y_B(R_0,\theta_0)$ and the global value $\langle\vert Y_B\vert\rangle$, where the former depends on the amount of CP violation in the initial VEVs, namely $\{R_0, \theta_0\}$ obtained from the stochastic inflation, and the latter is closely related to the real observable that has been averaged over the probability distribution in both the $R$ and $\theta$ directions. Due to the stochasticity of the $\theta$ distribution under mild $U(1)$ symmetry breaking, the global value $\langle\vert Y_B\vert\rangle$ is not restricted by the correlation length $R_c$ of the VEVs in the radial direction $R$,
\footnote{By ``global value'' we mean the ensemble average of the variance $\langle Y_B^2\rangle$, which is referred as the ``local value'' in the Appendix~A of \cite{Hook:2015foa}.}
where $R_c$ is characterized by the scalar mass $m$ \cite{Starobinsky:1994bd}.
We explore the isocurvature constraint for the baryogenesis with stochastic initial conditions in the mass range $0 < m/H_\ast \lesssim 1$, as shown in Section~\ref{Sec.isocurvature}.
Finally, a summary discussion is given in Section~\ref{Sec.summary}.

\section{Stochastic inflation with CP invariance}\label{Sec.stochastic_inflation}
The accumulation of long-wavelength fluctuations in a scalar field during inflation gives rise to a coherent condensate proportional to the inflationary Hubble parameter $H_\ast$. Since $H_\ast $ could be as high as $10^{13 - 14}$ GeV, the scalar field in general gain a sizable vacuum expectation value (VEV) during inflation, depending on the exact form of its potential. These large VEVs provide desirable initial conditions for the baryon production via the relaxation of a scalar condensate in the post-inflationary Universe. The Affleck-Dine baryogenesis \cite{Affleck:1984fy} is a typical scenario of this class. (See also \cite{Lloyd-Stubbs:2020sed,Hertzberg:2013mba,Takeda:2014eoa,Cline:2019fxx,Cline:2020mdt,Lin:2020lmr,Babichev:2018sia,Bamba:2016vjs} for the Affleck-Dine baryogenesis led by inflaton or heavy fields during inflation.)

One of the attractive features of baryogenesis from the inflationary condensate of a charged scalar is that the scenario can be constructed out of a CP invariant theory \cite{Hook:2015foa,Dolgov:1991fr,Linde:1985gh}. The basic idea is that even if the probability average of the baryon number density vanishes, namely $\langle n_B\rangle = 0$, as a consequence of the CP invariance, the expected variance $\langle n_B^2\rangle$ is non-zero due to the stochastic distribution of the quantum fluctuations during inflation that give rises to spontaneous CP violation. Baryogenesis is realized in a local patch of the Universe via renormalizable interactions that break the baryon number \cite{Hook:2015foa}. 

The condensation of scalar fields in de Sitter spacetime can be computed by the so-called ``stochastic inflation'' formalism, and the derivation of the basic equations applied in this work can be found in \cite{Starobinsky:1994bd}.
Previous efforts on the multi-scalar generalization of stochastic inflation are given in \cite{Mollerach:1990zf,Assadullahi:2016gkk,Vennin:2016wnk}, and see \cite{Adshead:2020ijf} for the cases with continuous symmetries.
In order to apply stochastic inflation for baryogenesis, we require an extended multi-field formalism with a mild-breaking of the continuous symmetry.

As a simple model, we consider a complex scalar field $\phi$ with subdominant density during inflation, and the Lagrangian of $\phi$ is given by
\begin{align}\label{AD_Lagrangian}
\mathcal{L} = \vert \partial_\mu\phi\vert^2 - m^2 \vert\phi\vert^2 + \delta\mathcal{L},
\end{align}
where $\delta\mathcal{L}$ can include all possible higher-order interactions. In the limit of $\delta\mathcal{L}\rightarrow 0$, the theory has a conserved current
\begin{align}\label{baryon_current}
j_B^\mu = i \left(\phi^\ast \partial^\mu \phi - \phi \partial^\mu\phi^\ast\right),
\end{align}
which is referred to as baryon number. The mass scale $m$ may be new physics associated with the Hubble parameter of inflation $H_\ast$, thus the mass term dominates the field potential so that there is no special ``flat direction.'' In the standard paradigm of inflation, $\phi$ reaches the equilibrium state as a massive field with the expectation value $\langle\phi\rangle =0$. 
When $\delta\mathcal{L} = 0$, the baryon number density $n_B = j_B^0 = 0$.

The presence of $\delta\mathcal{L}$ breaks the phase symmetry $\phi \rightarrow e^{i \alpha}\phi$ of the theory and thus violates the baryon number. 
The baryon violating interactions may be led by a set of quartic couplings as
\begin{align}\label{interaction general}
\delta\mathcal{L} \supset \sum_{m = 0}^{2} \lambda_m \left(\phi^{4 -m }\phi^{\ast\,m} + c. c.\right),
\end{align}
where $\lambda_m$ are small coupling constants ($\lambda_m \ll 1$) of the same order.
To make a concrete illustration of the basic idea in this scenario, we firstly set $\lambda_2 = \lambda_0 = 0$ and focus on a specific quartic coupling of the form
\begin{align}\label{interaction toy model}
\delta\mathcal{L} = \lambda_1 \left(\phi^3\phi^\ast + \phi^{\ast 3}\phi\right).
\end{align}
After clarifying the mechanism for baryogenesis we can easily extend our conclusion to all other types of interactions. We restrict ourselves to the case with $\Im[\lambda_m] = 0$ so that the theory is CP invariant.
In a spatially homogeneous background, the equation of motion in an expanding Universe reads
\begin{align}\label{eom:phi}
\ddot{\phi} + 3H\dot{\phi} +m^2 \phi = \lambda_1 \left(\phi^3 + 3\phi^{\ast 2}\phi\right).
\end{align}
We will perform two kinds of decomposition of the charged scalar as
\begin{align}\label{field decomposition}
\phi =\frac{1}{\sqrt{2}}\left( \phi_R + i \phi_I \right) = \frac{1}{\sqrt{2}} R \, e^{i \theta},
\end{align}
where the first representation is the standard decomposition commonly used in previous works \cite{Affleck:1984fy,Dolgov:1991fr,Hook:2015foa}. 
For some specific baryon violating interactions, such as \eqref{interaction toy model}, the first representation can lead to decoupled equations of motion for $\phi_{R}$ and $\phi_I$. In such a case, the baryon asymmetry can be readily estimated by the unique VEV computed in the decoupled system. As shown in the following sections, the second representation in terms of $R$, $\theta$ is more convenient for the discussion of baryon asymmetry with general interactions where the two scalar modes are mixed in the equations of motion.

Following the standard procedure of the stochastic inflation \cite{Starobinsky:1994bd}, one separates the long and short wavelength modes of $\phi$ and treats the short-wavelength modes as stochastic noises to the slow-roll equation of the long-wavelength modes. The probability distribution function (PDF) $\rho[\phi]$ of the charged scalar then obeys the multi-field Fokker-Planck equation \cite{Adshead:2020ijf,Starobinsky:1994bd}:
\begin{align}\label{FP_general}
\frac{\partial}{\partial t} \rho [\phi] = \frac{1}{3H} \vec{\nabla}\cdot \left(\rho [\phi]\vec{\nabla}V(\phi)\right) +\frac{H^3}{8\pi^2} \nabla^2\rho [\phi] = \vec{\nabla}\cdot\vec{J},
\end{align}
where the field derivative $\vec{\nabla}$ is defined with respect to the field space of the chosen representation \eqref{field decomposition}. For the standard representation $\vec{\nabla}\equiv \hat{\partial}_{\phi_{R}} + \hat{\partial}_{\phi_{I}} $ and for the polar representation $\vec{\nabla} \equiv \hat{\partial}_R + R^{-1}\hat{\partial}_\theta$. $\vec{J}$ is the probability current given by
\begin{align}\label{current general}
\vec{J} =  \frac{1}{3H} \rho \vec{\nabla} V +\frac{H^3}{8\pi^2} \vec{\nabla}\rho.
\end{align}
$V(\phi)$ is the effective potential in the classical equation of motion of $\phi$. The existence of an equilibrium state corresponds to a constant solution that satisfies $\partial_t \rho = \vec{\nabla} \cdot\vec{J} = 0$ at $t\rightarrow \infty$. The equilibrium state is a good approximation if the relaxation time scale $\tau_{\rm rex} \sim \Lambda_0^{-1}$ is much smaller than the duration of inflation, where $\Lambda_0$ is the lowest non-vanished eigenvalue among the eigenstates of $\rho[\phi]$ \cite{Starobinsky:1994bd} (see also Appendix~\ref{Appendix A}). 

\subsection{The standard representation}

We are mostly interested in the cases where the right-hand-side of \eqref{eom:phi} is perturbatively small, where the system exhibits two weakly coupled scalar degrees of freedom. Adopting the first decomposition in \eqref{field decomposition}, the classical equations of motion \eqref{eom:phi} are
\begin{align}\label{eom:phi_R full}
\ddot{\phi}_R + 3H\dot{\phi}_R +m^2 \phi_R &= \lambda \phi_R^3, \\\label{eom:phi_I full}
\ddot{\phi}_I + 3H\dot{\phi}_I +m^2 \phi_I &= -\lambda \phi_I^3,
\end{align}
where we use $\lambda = 4 \lambda_1$ for convenience.
Note that for the specific interaction \eqref{interaction toy model} the two scalars are fully decoupled in this representation.

Since $V(\phi_R, \phi_I) = V_R(\phi_R) + V_I(\phi_I)$, one can solve the Fokker-Planck equation by using the separation of variables $\rho[\phi_R, \phi_I] = \rho_R[\phi_R]\rho_I[\phi_I]$. It is easy to see that \eqref{FP_general} becomes a pair of decoupled equations for $\phi_R$ and $\phi_I$. In the limit of $\lambda \rightarrow 0$, the expectation values are $\langle\phi_R^2\rangle_0 = \langle\phi_I^2\rangle_0 = 3H_\ast^4/(8\pi^2m^2)$ readily given by the single-field cases \cite{Starobinsky:1994bd}. These are our expectation values at the zeroth order. 

To find out the corrections due to a non-zero $\lambda \ll m^2/\langle\phi^2\rangle_0$, we derive 
the effective potentials from the classical equations of motion as
\begin{align}\label{potential_leading_order}
V_R(\phi_R) = \frac{1}{2}m^2\phi_R^2 - \frac{\lambda}{4}\phi_R^4, \quad \text{and} \quad V_I(\phi_I) = \frac{1}{2}m^2\phi_I^2 + \frac{\lambda}{4}\phi_I^4.
\end{align}
For $\lambda > 0$ ($\lambda < 0$), $V_R$ ($V_I$) is meta-stable, therefore one has to take into account the condition that the escape rate from inside the potential is highly suppressed during inflation. We provide the detailed computation of $\langle\phi_{R, I}^2\rangle $ in Appendix~\ref{Appendix A}. The results are
\begin{align}\label{eq:phi_condesate}
\langle\phi_R^2\rangle = \frac{3 H_\ast^4}{8\pi^2 m^2}\left(1+ \frac{9\lambda H_\ast^4}{8\pi^2 m^4}\right), \\\nonumber
\langle\phi_I^2\rangle = \frac{3 H_\ast^4}{8\pi^2 m^2}\left(1 - \frac{9\lambda H_\ast^4}{8\pi^2 m^4}\right),
\end{align} 
for $\lambda \ll m^4/H_\ast^4$.
These solutions are the expectation values of the initial conditions for the relaxation of $\phi_{R, I}$ after the end of inflation. Note that in this representation the baryon number density is
\begin{align}\label{B number decoupling}
n_B = \phi_R \dot{\phi}_I - \phi_I\dot{\phi}_R.
\end{align}
Since equilibrium states are time-independent until the end of inflation, we have the initial condition $\dot{\phi}_R = \dot{\phi}_I = 0$ for the relaxation of the field condensate. 
One can see that the presence of a non-zero VEV for $\phi_{I}$ in the initial condition \eqref{eq:phi_condesate} generally leads to a different dynamics for $\phi$ and $\phi^\ast$, which implies a spontaneous breaking of the CP invariance. As a result, a non-zero baryon number $n_B$ can be generated after $\phi_{R}$ and $\phi_{I}$ start in motion.

\subsection{The polar representation}
The second representation $\phi = R \,e^{i\theta}/\sqrt{2}$ in \eqref{field decomposition} provides a more systematic understanding on the generation of baryon number with general baryon violating interactions (see also \cite{vonHarling:2012yn,Bell:2011tn,Bamba:2018bwl,Barrie:2020hiu,Dine:1995kz}) since the decoupling of the equations of motion may not be always possible. 
Given that $\delta\mathcal{L}$ breaks the $U(1)_\phi $ symmetry, we have a mixed PDF $\rho = \rho[R, \theta]$ for the two random fields to be obtained from \eqref{FP_general}. Note that in the polar representation, $\vec{\nabla} = \hat{\partial}_R + R^{-1}\hat{\partial}_\theta$ and $\nabla^2 = R^{-1}\partial_R(R\partial_R) + R^{-2} \partial^2_\theta$.

If we are only interested in the PDF with respect to the lowest eigenvalue, it is useful to obtain the solution via the probability current $\vec{J}$ \cite{Risken:1984}.
The condition for the existence of a (quasi-)equilibrium state indicates that $\vec{J}$ is a constant and thus $\vec{J}\rightarrow 0$ at the boundary of the potential leads to the solution of \eqref{current general} as
\begin{align}\label{PDF_polar}
\rho[R, \theta] = \exp\left[-\frac{8\pi^2}{3H_\ast^4}V(R,\theta)\right]/\Sigma,
\end{align}
where the effective potential including the specific interaction \eqref{interaction toy model} reads
\begin{align}\label{potential_polar}
V(R, \theta) = \frac{1}{2}m^2 R^2 + \frac{\lambda}{2}R^4 \cos(2\theta).
\end{align}
In terms of the dimensionless parameters $r = R/H_\ast$ and $r_0 \equiv (3H_\ast^2/(4\pi^4m^2))^{1/2}$, the normalization constant $\Sigma$ is  defined as
\begin{align}
\Sigma = \int_{0}^{2\pi}\int_{0}^{r_\ast}\exp\left[-\frac{r^2}{r_0^2} - \frac{4}{3} \pi^2\lambda r^4 \cos(2\theta) \right] r dr d\theta,
\end{align}
which ensures $\int_{0}^{2\pi}\int_{0}^{r_\ast} \rho[r,\theta] r dr d\theta = 1$. The upper limit $r_\ast$ is in general $\theta$-dependent since $\lambda \cos(2\theta)$ changes sign every $n\pi/4$. For $\lambda \cos(2\theta) < 0$, $V(R, \theta)$ is meta-stable so that $r_\ast$ shall satisfy the condition for the stability of the quasi-equilibrium state. As shown in Appendix~\ref{Appendix A}, this gives a conservative upper bound $r_\ast \leq m/(\sqrt{\vert\lambda\cos(2\theta)\vert}H_\ast)$. Of course we expect higher-order corrections or new physics that stabilizes the potential to be important at a value of $r$ which does not depend on $\theta$. 

On the other hand, if $r_\ast$ is chosen such that $\frac{4}{3}\pi^2 \lambda  r_\ast^4 \ll 1$, we may use the approximation to obtain
\begin{align}
\Sigma &\simeq \int_{0}^{2\pi}\int_{0}^{r_\ast} e^{-r^2/r_0^2} \left(1- \frac{4}{3} \pi^2 \lambda \cos(2\theta) r^4\right) r dr d\theta \\
&= \pi r_0^2 \left(1-  e^{-r_\ast^2/r_0^2}\right).
\end{align}
In this regime, the variance can be computed analytically
\begin{align}
\left\langle r^2 \right\rangle &\simeq \frac{1}{\Sigma} 
\int_{0}^{2\pi}\int_{0}^{r_\ast} e^{-r^2/r_0^2} \left(1- \frac{4}{3} \pi^2 \lambda \cos(2\theta) r^4\right) r^3 dr d\theta, \\
&= r_0^2 \left(1+\frac{r_\ast^2}{r_0^2}\frac{1}{e^{r_\ast^2/r_0^2} -1} \right).
\end{align} 
Note that $r_0^2 = 3H_\ast^2 /(4\pi^2 m^2)$ is nothing but the expectation value $\left\langle r^2 \right\rangle$ in the case with $\lambda  =0 $, and $\left\langle r^2 \right\rangle \rightarrow r_0^2$ in the limit of $r_\ast^2 /  r_0^2 \rightarrow \infty$. We require $r_\ast^2 /  r_0^2 \gg 1$ in the following discussion. The correction to $\left\langle r^2 \right\rangle$ due to the baryon violating interaction enters from $\mathcal{O}(\lambda^2)$.
It is also easy to check that $\langle \cos^2\theta\rangle = 1/2 -\frac{2}{3}\pi^2\lambda r_0^4$ following the linear expansion of the $\lambda$-correction for the PDF $\rho[r, \theta]$.
The angular distribution is nearly homogeneous as the baryon-violating coupling is small.

Remarkably, the VEVs $\phi_{R0} \equiv \pm \langle \phi_{R}^2\rangle^{1/2}$ and $\phi_{I0} \equiv\pm \langle \phi_{I}^2\rangle^{1/2}$ given by \eqref{eq:phi_condesate} with the leading order correction of $\mathcal{O}(\lambda)$ correspond to the specific VEVs in the polar representation as $R_0^2 = \phi_{R0}^2 + \phi_{I0}^2 \approx r_0^2 H_\ast^2$ and $\theta_0 = \tan^{-1}(\phi_{I0}/\phi_{R0}) = n \pi/4 + \mathcal{O}(\lambda)$. In the limit of $\lambda \rightarrow 0$ it appears $\theta_0 \rightarrow n \pi/4$, yet the restored $U_\phi(1)$ symmetry is equivalent to imposing a shift symmetry in the angular direction ($\theta_0 \rightarrow \theta_0 +c$). The presence of a non-zero $\lambda$ leads to a mild deform on the probability distribution of the angular VEV but an important difference on the resulting baryon asymmetry via different values of $\theta_0$. We will clarify this point in the next section.

The above expectation values will be used as initial conditions for solving the relaxation dynamics in the post-inflationary epoch, where the classical equations of motion in the polar representation are
\begin{align}\label{eom: polar_R}
\ddot{R} + 3H\dot{R} + \left(m^2 - \dot{\theta}^2\right) R = \lambda R^3 \cos(2\theta), \\ \label{eom: polar_theta}
\ddot{\theta} + \left(3H + 2 \frac{\dot{R}}{R}\right) \dot{\theta} = -\frac{\lambda}{2} R^2 \sin(2\theta).
\end{align}
The baryon number density is given by
\begin{align}\label{B number polar}
n_B =  R^2 \dot{\theta}.
\end{align}
Since in this approximation the equilibrium state is constant in time, we have the initial conditions $\dot{R} = \dot{\theta} = 0$. Thus the baryon production is equivalent to the generation of a non-zero $\dot{\theta}$ in the post-inflationary relaxation of the field condensate.

\section{Relaxation baryogenesis and spontaneous CP violation}\label{Sec.baryogenesis}
As inflation ends, the coherent field condensate is no longer in equilibrium with the cosmic expansion so that it starts to relax towards the potential minimum. 
The baryon production via the relaxation of a coherent scalar condensate is familiar in the Affleck-Dine scenario \cite{Affleck:1984fy}.
To compute the baryon production for general interactions that may lead to a mixed equations of motion, we adopt the formalism based on the polar representation. 
In pioneering works \cite{Dolgov:1991fr,Hook:2015foa}, the baryon asymmetry are usually estimated by the standard representation with $\vert\phi_{R0}\vert \approx \vert \phi_{I0} \vert $, which in fact corresponds to the specific angles $\theta_0 \approx n\pi/ 4$ in the polar representation.
We show that the baryon asymmetry led by $\theta_0 \approx n\pi/ 4$ is indeed a good approximation for specific interactions, such as \eqref{interaction toy model}, but in general the angle of dominant baryon asymmetry can be different with other kinds of interactions. 
Having worked out the probability distribution function via stochastic inflation, in this section we address the expected baryon asymmetry from reheating to the beginning of radiation domination.

\begin{figure}
	\begin{center}
		\includegraphics[width=7cm]{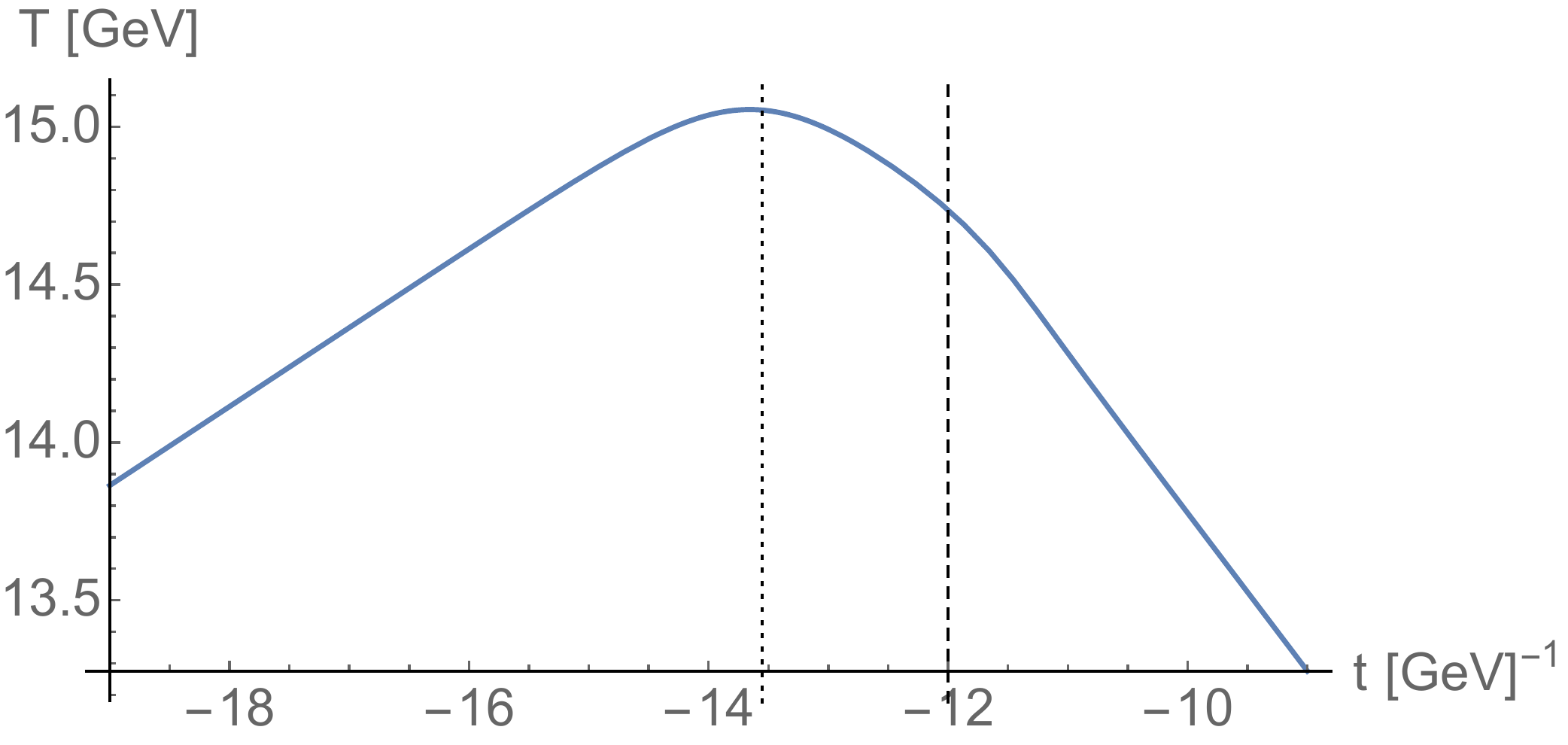}
	\end{center}
	\caption{The reheating temperature $T$ as a function of time $t$. The dotted line is the time for the maximal temperature and the dashed line is the starting time for the radiation domination. \label{fig:temperature}}
\end{figure}

\smallskip
\noindent
\textbf{Reheating background.}
For the background dynamics we consider the typical paradigm where inflation is followed by a matter dominated epoch due to the coherent oscillation of the inflaton field (preheating). During preheating the decay of inflaton into radiation is characterized by the decay width $\Gamma_I$. With the initial energy density of the inflaton being $\rho_I(t_0) = 3M_P^2H_\ast^2 = \Lambda_I^4$, the reheating of the Universe is described by
\begin{align}\label{reheating background}
\dot{\rho}_I + 3H \rho_I &= - \Gamma_I \rho_I, \\
\dot{\rho}_R + 4H \rho_R &= \Gamma_I \rho_I,
\end{align}
where $\rho_I$ is the smoothed energy density of the inflaton and $\rho_R$ is the radiation energy density. The density of the $\phi$ field is always negligible so that the Hubble parameter is $H^2 = (\rho_I +\rho_R)/(3M_P^2)$. This gives the analytic solution
\begin{align}
\rho_I(t) = \Lambda_I^4 a^{-3}(t)e^{-\Gamma_I (t - t_0)},
\end{align}
where $a(t_0) = 1$ is used. $\rho_R$ can be solved numerically with the analytic solution of $\rho_I$. One can approximate the temperature as a function of time via
\begin{align}
T(t) = \left(\frac{30}{\pi^2 g_\ast} \rho_R(t)\right)^{1/4},
\end{align}
where $g_\ast = 106.75$ is the relativistic degrees of freedom at a temperature higher than $300$ GeV. The maximal temperature of reheating $T_{\rm max} = T(t_{\rm max})$ occurs around the time $t_{\rm max} \approx 2/(3H_\ast)$. The time for the beginning of the radiation domination is $t_R \approx 1/\Gamma_I$. In Figure~\ref{fig:temperature} we give an example of the temperature evolution with $\Lambda_I = 10^{16}$ GeV and $\Gamma_I = 10^{12}$ GeV. This initial energy density corresponds to $H_\ast = 2.37 \times 10^{13}$ GeV.

\begin{figure}
	\begin{center}
		\includegraphics[width=8cm]{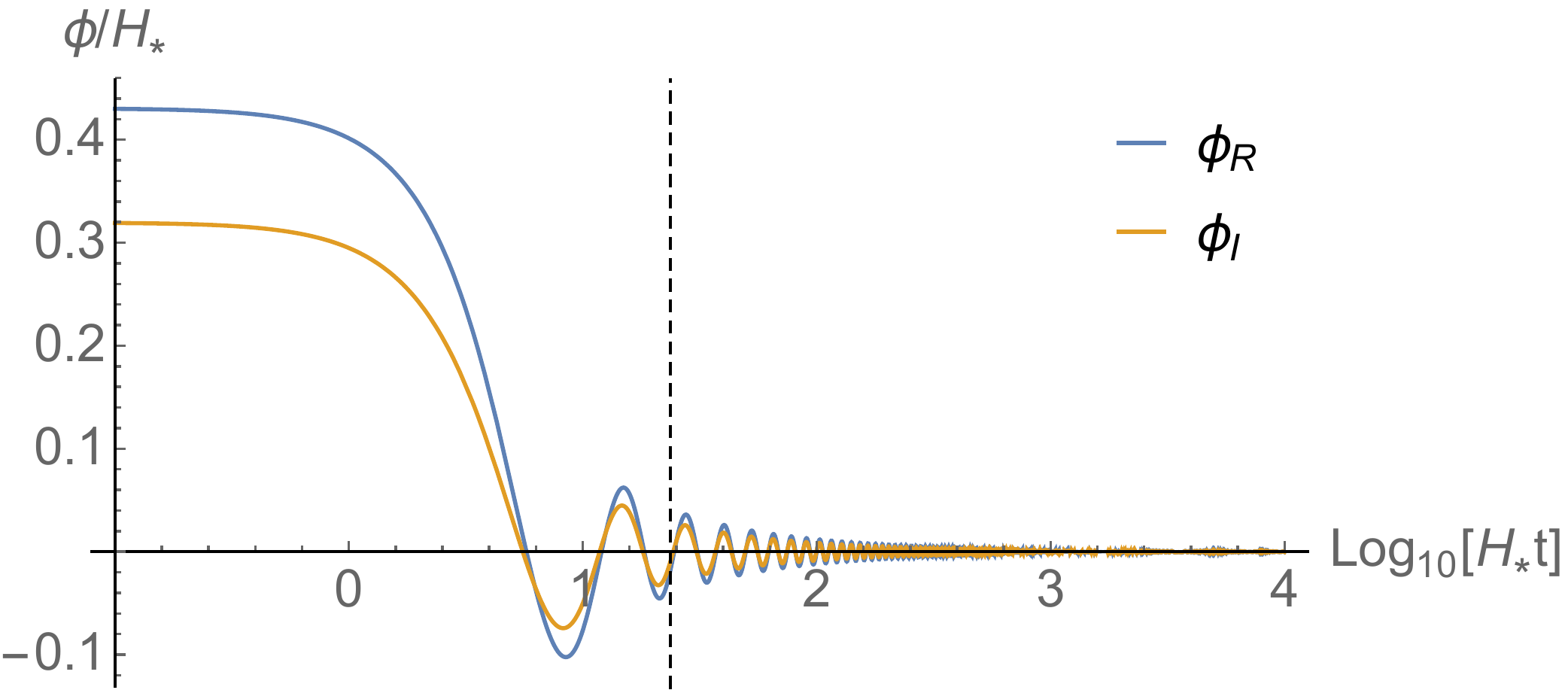}
	\end{center}
	\caption{Relaxation of the field condensate with $m = 0.5 H_\ast$ and $\lambda =0.12$ from the decoupling initial condition \eqref{eq:phi_condesate}.\label{fig:phi_VEV}}
\end{figure}

\begin{figure}
	\begin{center}
		\includegraphics[width=8cm]{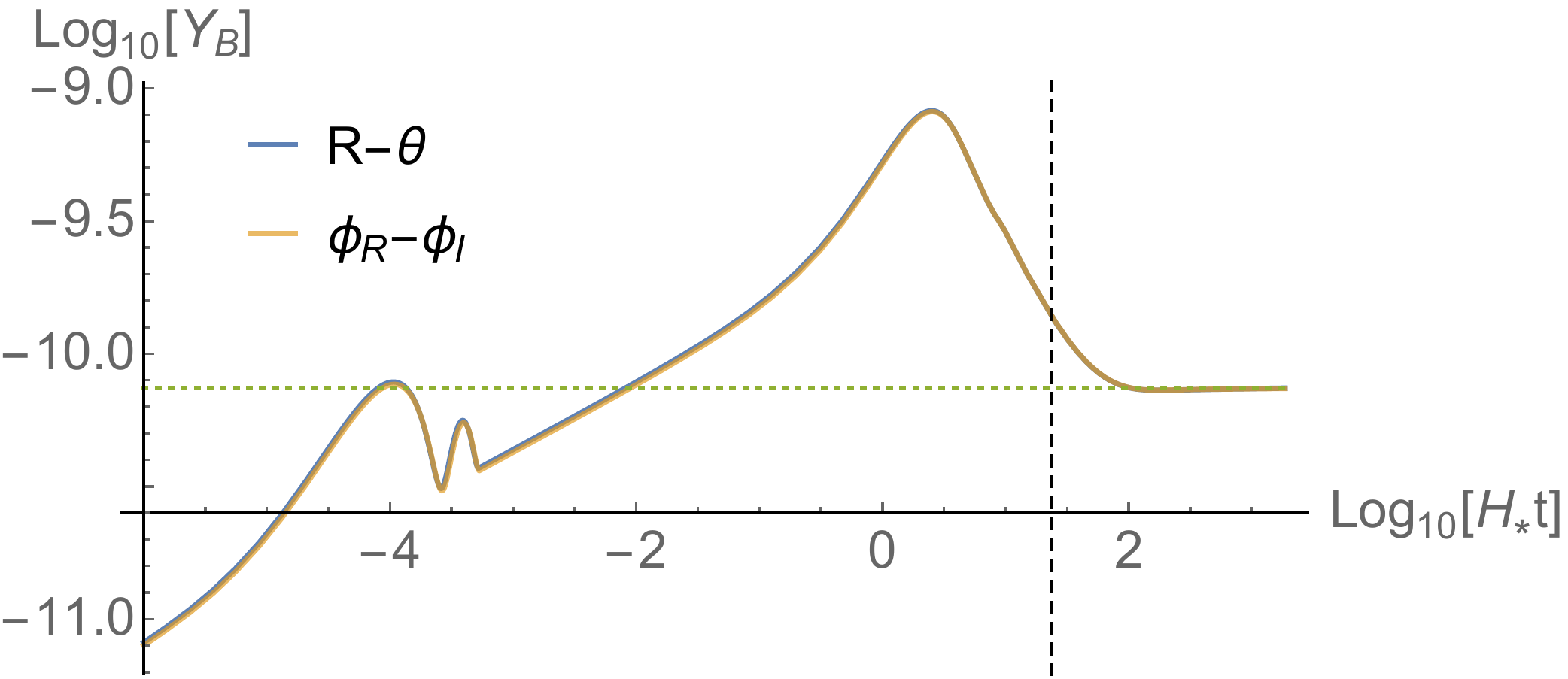} 
	\end{center}
	\caption{The baryon number density as a function of time with $m = 0.5 H_\ast$ and $\lambda =0.12$. The parameters are $\Lambda_I=1\times10^{16}\,\text{GeV}$ and $\Gamma_{I}=10^{12}\,\text{GeV}$. The vertical dashed line is the time for the beginning of radiation domination. The initial conditions for $\{\phi_{R},\phi_{I}\}$ are given by \eqref{eq:phi_condesate} and $\{R_0 = r_0H_\ast, \theta_0 = \pi/4\}$ are used. \label{fig:Y_B}}
\end{figure}

We can solve the relaxation dynamics of the charged scalar, with the background expansion led by \eqref{reheating background}, via the decoupling representation \eqref{eom:phi_R full} and \eqref{eom:phi_I full}, or the polar representation \eqref{eom: polar_R} and \eqref{eom: polar_theta}. As an example we show in Figure~\ref{fig:phi_VEV} the evolution of $\{\phi_{R}, \phi_I \}$ 
with $\lambda = 0.12$, $m = 0.5 H_\ast$ based on the background given by Figure~\ref{fig:temperature}. The initial conditions for $\phi_{R, I}$ are given by \eqref{eq:phi_condesate}. 
The final baryon asymmetry diluted by the entropy production $s = 2\pi^2 g_\ast T^3/45$ during reheating is 
\begin{align}\label{B_asymmetry_definition}
Y_B(t) = \frac{n_B(t)}{s(t)} = \frac{45}{2\pi^2 g_\ast}\frac{n_B(t)}{T^3(t)},
\end{align}
where $n_B(t)$ is evaluated by \eqref{B number decoupling}. As shown in Figure~\ref{fig:Y_B}, the maximal baryon asymmetry is generated around the onset of the field oscillation at $t_{\rm osc} = H^{-1}_{\rm osc} \simeq m^{-1}$.

\smallskip
\noindent
\textbf{Spontaneous CP violation.}
In the case with mild $U(1)_\phi$ symmetry breaking, the angular mode (namely $\theta$ in the polar representation) is nearly a massless field whose VEV remains a stochastic variable at the end of inflation. As long as the field relaxation starts from an initial value $\theta_0$ with non-zero angular gradient (namely $\partial_\theta V \neq 0$ at $\theta_0$), a non-zero angular velocity $\dot{\theta}$ will convert from the potential energy and thus result in non-zero baryon number in the local patch of the Universe. 
\footnote{The size of the local Universe with the same baryon number is characterized by the correlation length of the inflationary condensate led by $R_0$ \cite{Hook:2015foa,Starobinsky:1994bd}. However, the theory has no predictability on the local baryon number, due to the stochasticity of the initial VEV $\theta_0$.} 
This is the consequence of the spontaneous CP violation led by the initial condition $\theta_0$. 
Note that the interaction \eqref{interaction toy model} is invariant under the rotation $\phi \rightarrow \phi \,e^{i n \pi /2}$ where $ \theta = n \pi /2$ are special angles that preserves CP. Similarly, the $\lambda_0$ interaction in \eqref{interaction general} preserves CP at $\theta = n \pi/ 4$.

\begin{figure}
	\begin{center}
		\includegraphics[width=9.2cm]{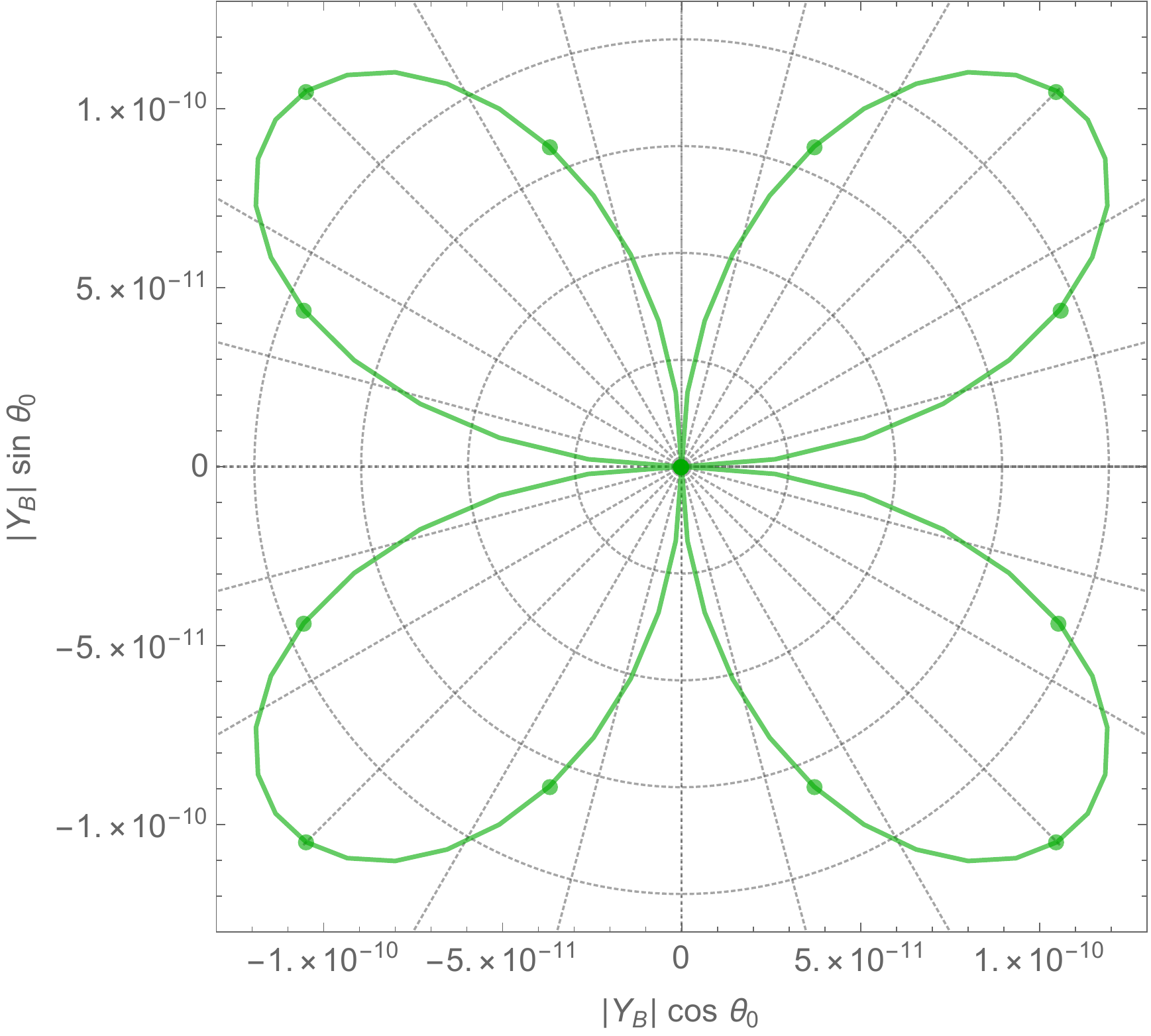}
	\end{center}
	\caption{A scan of the final baryon asymmetry $\vert Y_B\vert$ as a radial function of $\theta_0 \in [ 0, 2\pi ]$ at the fixed value $R_0 = \sqrt{3H_\ast^4/(4\pi^2 m^2)} = r_0H_\ast$, where $Y_B = Y_B(R_0, \theta_0)$ is the numerical result at reheating complete with respect to initial conditions $\{R_0, \theta_0\}$.
		We use $m = 0.5 H_\ast$. 
		The coupling parameters are $\lambda_0 = 0$ and $\lambda = 4 \lambda_1 = 0.12$ (green-solid line). The dots correspond to $\theta_0 = n\pi/8$. \label{fig:angular_scan}}
\end{figure}

The baryon production (from a local initial condition $\{R_0, \theta_0\}$) in this scenario can be schematically divided into two stages. The first stage is the creation of a net baryon number when $t < t_{\rm osc} \simeq m^{-1}$. To calculate the generation of baryon number it is more convenient to use the polar representation and focus on the dynamics of $\dot{\theta}$ via the baryon violation source on the right-hand side of \eqref{eom: polar_theta}.
Since $H \gtrsim m$ in this stage, the friction due to the expansion of the Universe implies that $R$ is slowly decreasing from the initial condition $R_0$ and thus $H \gg \dot{R}/R$ so that $\theta = \theta_0$ is essentially a constant during this stage. 
At the onset of the field oscillation at $t = t_{\rm osc} \simeq m^{-1}$, (before $\theta$ receives a kick) $\theta$ acquires a small but non-zero angular velocity $\dot{\theta}_{\rm osc} \equiv \dot{\theta}(t_{\rm osc}) \approx \frac{\lambda}{6H_\ast}R_0^2\sin(2 \theta_0)$ for $\theta_0 \neq n \pi /2$.
As shown in Figure~\ref{fig:Y_B}, the baryon number density $n_B(t)$ reaches its peak value at $t \approx m^{-1}$ and this can be estimated as $\vert n_B \vert \approx \lambda R_0^4 \sin(2 \theta_0)/ (3H_\ast)$, which is a good approximation for $m > 0.5 H_\ast$.
We expect that the baryon number is mainly created at $\theta \approx  n \pi/4$ with $n = 1, 3, 5, 7$, where CP violation is maximal.
A scan of the $\theta_0$-dependence of the resulting asymmetry $Y_B(R_0, \theta_0)$ is given as the green-solid line (with $\lambda = 4 \lambda_1 = 0.12$) in Figure~\ref{fig:angular_scan} for the initial condition $R = R_0 = \sqrt{3H_\ast^4/(4\pi^2 m^2)}$. The maximal asymmetry exhibits a small deviation from $\theta_0 = n \pi/4$ due to the small baryon violating corrections.
Note that our definition gives $Y_B < 0$ for $\theta_0 = \pi/4$ or $5\pi/4$, and in the polar representation we use \eqref{B number polar} for $n_B$. 
Figure~\ref{fig:Y_B} shows that the baryon asymmetry $\vert Y_B \vert$ generated from the initial conditions $\{R_0 = r_0H_\ast, \theta_0 = \pi/4\}$ agrees with that from the initial condition \eqref{eq:phi_condesate} in the decoupling representation, where $R_0^2 = \phi_{R0}^2 +\phi_{I0}^2$.

\begin{figure}
	\begin{center}
		\includegraphics[width=6cm]{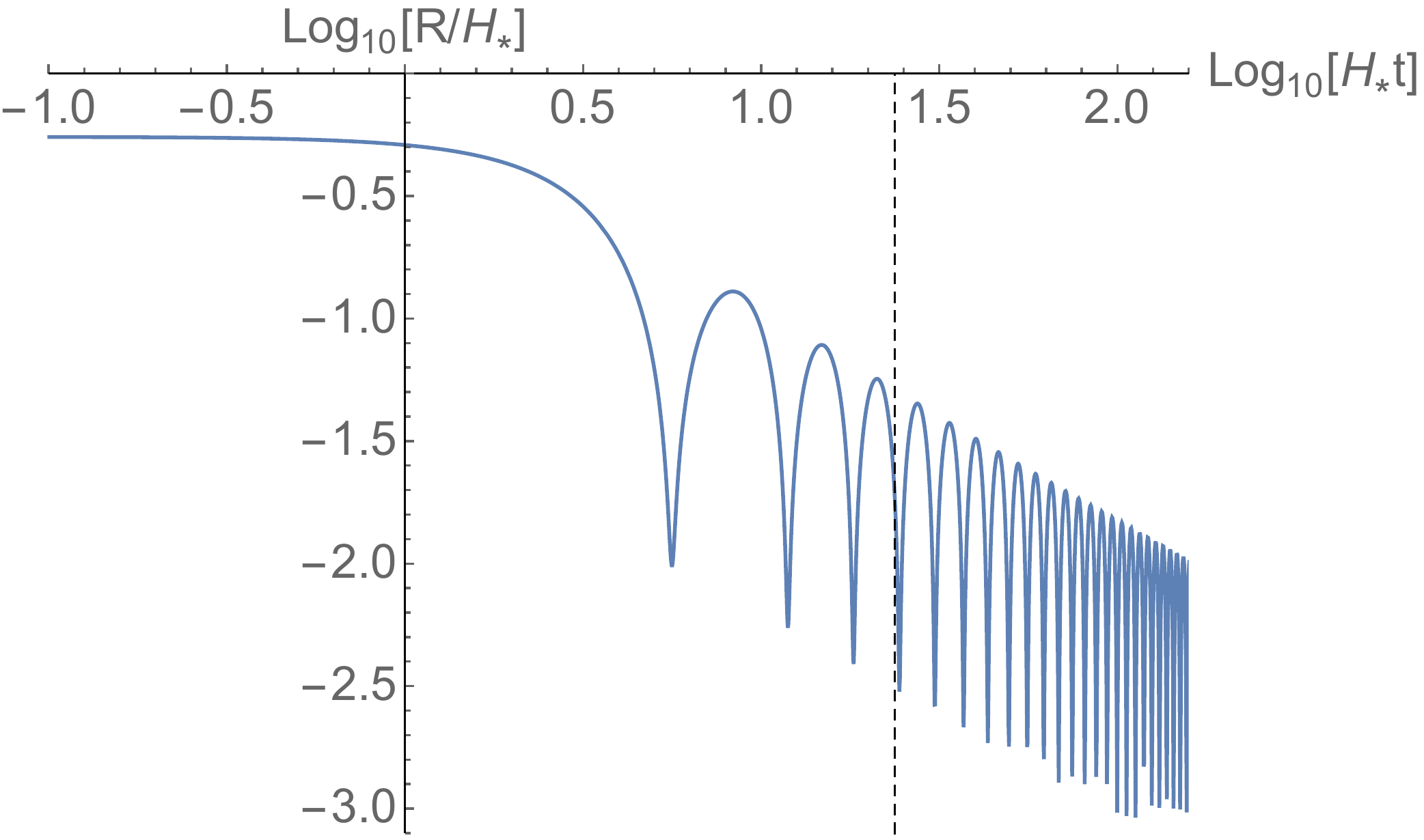} 
		\hfill
		\includegraphics[width=6cm]{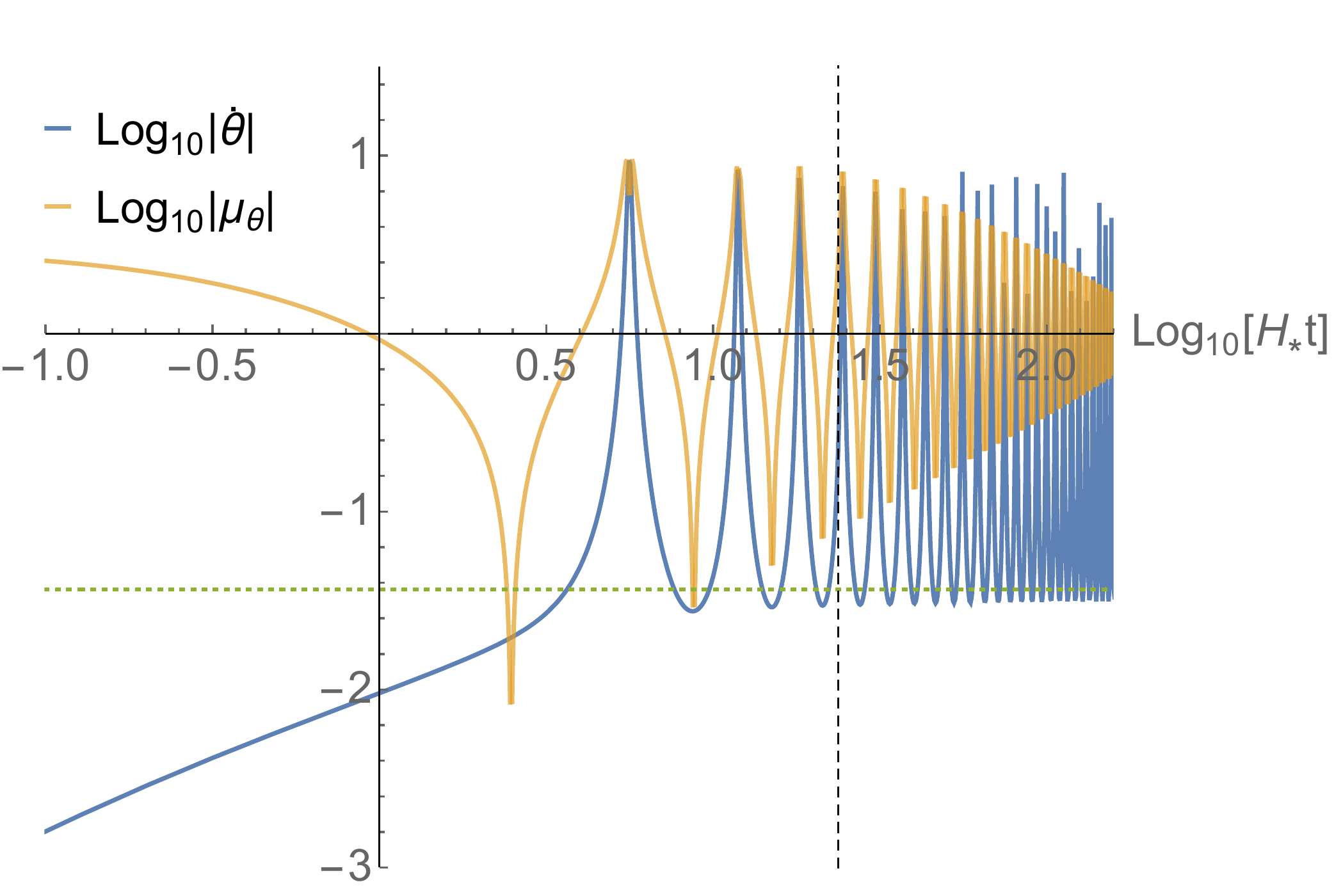} 
	\end{center}
	\caption{Relaxation of the field condensate in the polar representation with $m = 0.5 H_\ast$ and $\lambda =0.12$. The initial conditions are $R_0  = \sqrt{3H_\ast^4/(4\pi^2 m^2)} = r_0 H_\ast$ and $\theta_0 = \pi/4$. The green-dotted line is the analytic approximation $\dot{\theta}_R = \lambda R_0^2/(2m)$. The vertical dashed line is the time for the beginning of radiation domination. \label{fig:R}}
\end{figure}

The second stage is the dilution of baryon asymmetry from the onset of field oscillation to the complete of reheating for $t_{\rm osc} < t < t_R$.
As $R$ starts to oscillate across the origin, the angular velocity $\dot{\theta}$ evolves with sharp oscillations. Let us define $\mu_\theta \equiv 3H + 2\dot{R}/R$ and rewrite \eqref{eom: polar_theta} around $R \rightarrow 0$ as
\begin{align}
\ddot{\theta} + \mu_\theta \dot{\theta} \approx 0.
\end{align}
The solution takes the form of $\dot{\theta} \simeq \dot{\theta}_{\rm osc} e^{-\int \mu_\theta dt}$.
Since $\mu_\theta \approx 2\dot{R}/R$ when $R \rightarrow 0$, the sign of $\mu_\theta$ is the same as the sign of $\dot{R}$. As shown in Figure~\ref{fig:R}, $\dot{\theta}$ obtains a spiky enhancement each time when $\mu_\theta$ turns into negative values, and the most important enhancement of $\dot{\theta}$ is led by the first oscillation period where $\vert \mu_\theta\vert$ has the largest amplitude. Due to the damping of the oscillating amplitude of $R$, the exponential enhancement with $\mu_\theta < 0$ is not perfectly canceled out by the exponential decay with $\mu_\theta > 0$ and thus $\dot{\theta}$ is further lifted from $\dot{\theta}_{\rm osc} $ at the end of reheating. 
The final value $\dot{\theta}_R \approx \dot{\theta}_{\rm osc} H_\ast/m$ is a good approximation for $m \leq H_\ast$.
The approximation for the case with $m = 0.5 H_\ast$ and $\theta_0 = \pi/4$ (maximum CP violation) so that $\dot{\theta}_R \approx \lambda R_0^2 / (2m)$ is depicted by the green-dotted line in the right panel of Figure~\ref{fig:R}.
Note that the baryon density $n_B(t)$ is decreasing during this stage despite $\dot{\theta}$ is slightly enhanced, due the the rapid decrease of $R$.

An order of magnitude estimation of the final baryon asymmetry at the beginning of radiation domination can be
\begin{align}
\left\vert Y_B(t_R) \right\vert = \frac{\vert n_B(t_R) \vert }{s(t_R)} \sim  \frac{\tilde{\lambda}(\theta_0) R_0^2}{2 m} R^2(t_R) \frac{45}{2\pi^2 g_\ast}\frac{1}{T^3(t_R)},
\end{align}
where we denote $\tilde{\lambda}(\theta_0) \equiv \lambda \sin(2\theta_0)$ with respect to the initial VEV $\theta_0$. 
The amplitude at $t_R$ is approximated by the scaling of the background density $R^2(t_R) \simeq c R_0^2 H^2(t_R) /m^2$, assuming that $R = R_0$ is nearly unchanged until the onset of oscillation around $t = m^{-1}$. The numerical factor $c \sim \mathcal{O}(1)$ accounts for the relative enhancement of $R \sim a^{-3/2}$ in the transition from matter domination to radiation domination.
As a result, one finds 
\begin{align}\label{B_asymmetry_approx}
\left\vert Y_B(t_R) \right\vert \sim  c  \frac{\tilde{\lambda}(\theta_0) R_0^4}{2 m^3} \frac{H^2(t_R)}{s(t_R)} \sim c  \frac{\tilde{\lambda}(\theta_0) R_0^4}{2 m^3} \frac{T_R}{4 M_p^2},
\end{align}
for $m \leq H_\ast$ where $H(t_R) \sim \Gamma_{I}^{-1}$ and $T_R \sim (90 M_p^2 \Gamma_{I}^2/g_\ast\pi^2)^{1/4}$. In Figure~\ref{fig:Y_B}, we show the approximation \eqref{B_asymmetry_approx} as the green-dotted line with $c\approx 1.5$.
We note that the result \eqref{B_asymmetry_approx} essentially differs from the estimation in \cite{Hook:2015foa} by the dilution factor $H^2(t_R)/m^2$ due to the fact that we consider a non-instantaneous reheating led by inflaton oscillation.

\begin{figure}
	\begin{center}
		\includegraphics[width=7.5cm]{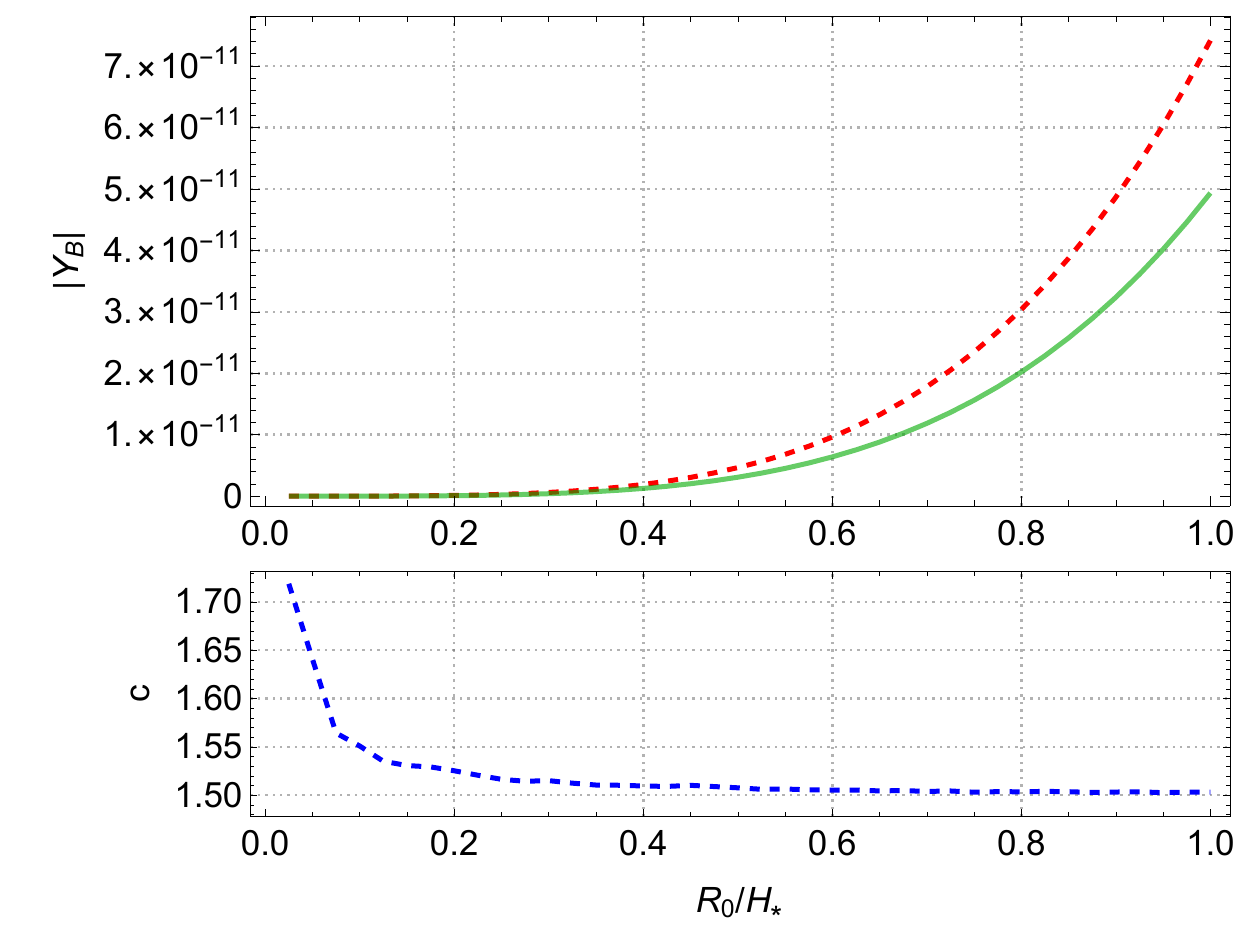} 
		\hfill
		\includegraphics[width=7.5cm]{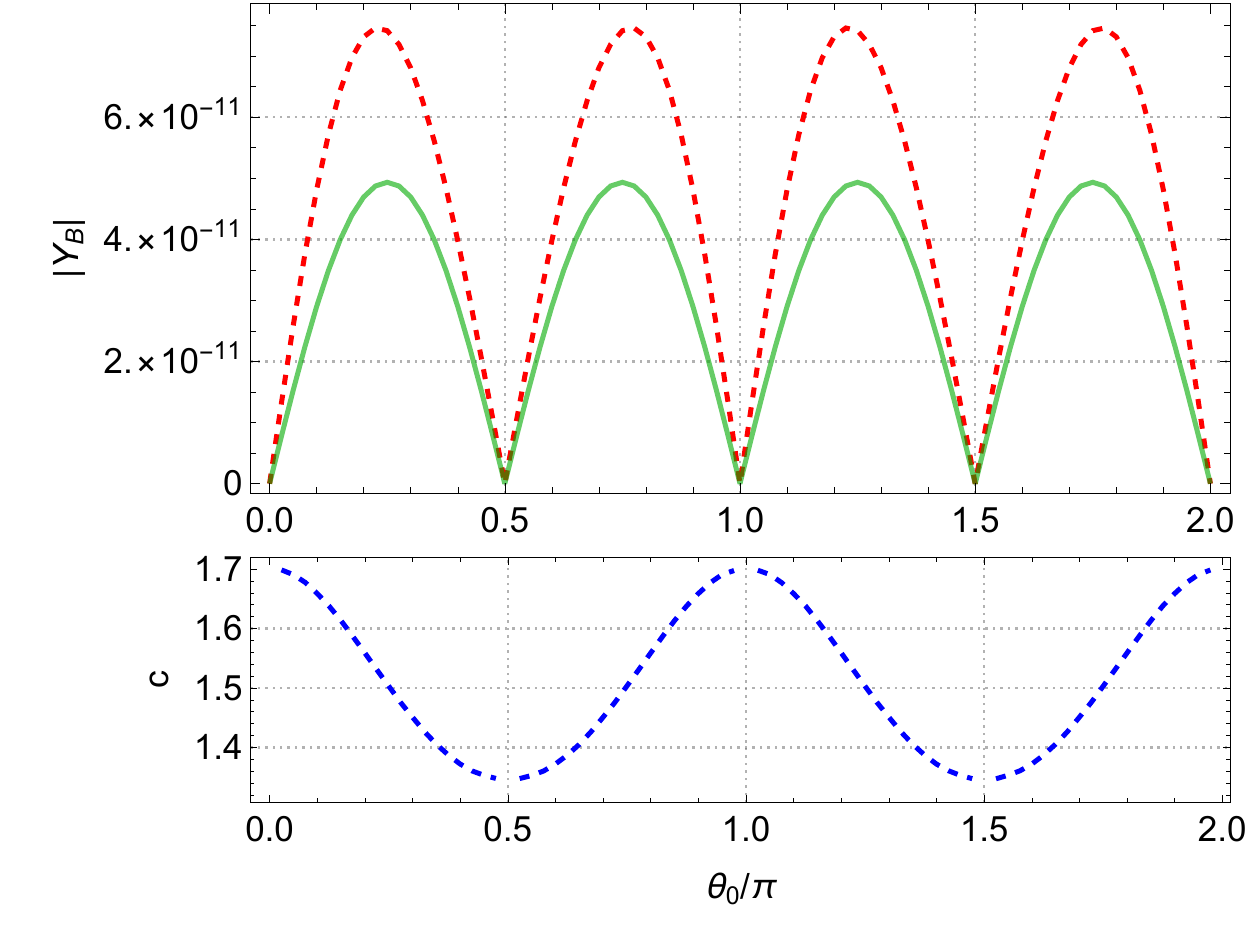} 
	\end{center}
	\caption{The numerical factor $c$ as the ratio between the numerical results of the final baryon asymmetry (red-dashed lines) and the analytic approximation \eqref{B_asymmetry_approx} with $c = 1$ (green-solid lines). The initial condition $\theta_0 = \pi/4$ is fixed in the left panel and $R_0  = \sqrt{3H_\ast^4/(4\pi^2 m^2)} = r_0 H_\ast$ is fixed in the right panel. Parameters with $m = 0.5 H_\ast$ and $\lambda =0.12$ are used in this figure. \label{fig:c_factor}}
\end{figure}

Let us test the numerical factor $c$ with respect to various initial conditions $R_0$ and $\theta_0$. To be more precise, we shall define $\vert Y_{B, \mathrm{num}} \vert = c\, \vert Y_{B, \mathrm{ana}} \vert$, where $Y_{B, \mathrm{num}}(R_0, \theta_0)$ is the final baryon asymmetry \eqref{B_asymmetry_definition} via numerical evaluation and $Y_{B, \mathrm{ana}}(R_0, \theta_0)$ is the analytic approximation at $t =t_R$ given by \eqref{B_asymmetry_approx} with $c =1$. In Figure~\ref{fig:c_factor} we compare numerical results with the analytic formula \eqref{B_asymmetry_approx} at fixed $R_0$ or $\theta_0$. We confirmed that the $c$ factor is $\mathcal{O}(1)$ 
and $c \approx 1.5$ around the maximal CP violation VEVs.

\smallskip
\noindent
\textbf{Baryon asymmetry.}
The baryon asymmetry \eqref{B_asymmetry_approx} indicates $Y_B(t_R) \propto \tilde{\lambda} \propto \sin(2\theta_0)$, which exhibits the property $Y_B(R_0, \theta_0) = - Y_B(R_0, \theta_0 + \pi/2)$ with respect to the initial VEVs $\{R_0, \theta_0\}$ (see also Figure~\ref{fig:angular_scan}).
This reflects the symmetry in the shape of the potential $V(R, \theta)$ of \eqref{potential_polar}. 
As a result, the global expectation value $\langle Y_B \rangle$ averaging over the probability distribution of all possible initial VEVs vanishes identically (due to the angular integration), which is the consequence of CP invariance in the original theory \eqref{interaction toy model}. 
However, we are more interested in the global variance of the final baryon asymmetry captured by the non-zero expectation value
\begin{align}\label{B_average_num}
\left\langle Y_B^2 \right\rangle  = \int_{0}^{2\pi} \int_{0}^{r_\ast} Y_B^2(r, \theta) \rho[r, \theta] r dr d\theta,
\end{align}
where we use the dimensionless parameter $r = R/H_\ast$ for convenience. Here $Y_B(r, \theta)$ stands for the numerical result at $t =t_R$ with $\{r, \theta\}$ the input initial VEVs, and $\rho[r, \theta]$ is the PDF of $\{r, \theta\}$ given by \eqref{PDF_polar} and \eqref{potential_polar}.
Our numerical results show that $\langle \vert Y_B \vert \rangle \equiv \langle Y_B^2 \rangle^{1/2}  = c_{\rm avg} \times \vert Y_{B, \mathrm{num}}(r_0, \theta_{\rm max}) \vert$ with $2.3 < c_{\rm avg} < 3.2$, depending on the choices of $r_\ast$. Note that we have defined $r_0 \equiv \langle r^2\rangle^{1/2}$ and $\theta_{\rm max} \approx n \pi /4$ for $n =1, 3, 5, 7$ are initial VEVs around maximal CP violation. 

Let us also examine the averaged asymmetry via the analytic approach. Replacing $Y_B(r, \theta)$ in \eqref{B_average_num} by $Y_{B, \mathrm{ana}}$ obtained in \eqref{B_asymmetry_approx}, we can derive (in the limit of $r_\ast^2/r_0^2 \gg 1$)
\begin{align}
\left\langle Y_B^2 \right\rangle_{\rm ana}  &= \int_{0}^{2\pi} \int_{0}^{r_\ast} Y_{B, \mathrm{ana}}^2(r, \theta) \rho[r, \theta] r dr d\theta, \\
&\simeq 12 \times Y_{B, \mathrm{ana}}^2(r_0, \theta_{\rm max}),
\end{align}
where the linear expansion of the $\lambda$ correction in the PDF $\rho[r, \theta]$ is used again. One can see that the $\theta$-dependence in the PDF drops out in the angular average due to the linear expansion of $\lambda$. The factor $12$ arises from the integration of $r^9e^{-r^2/r_0^2}$ with respect to $r$. Since $\sqrt{12} \approx 3.46$, the analytic result agrees with the numerical computation up to a $\mathcal{O}(1)$ difference.
The results from both the numerical and the analytical approaches indicate that the baryon asymmetry results from initial VEVs around the largest angular gradient $Y_B(r_0, \theta_{\rm max})$, or namely the maximal source term of \eqref{eom: polar_theta}, is approximately at the same order of the expected value $\langle \vert Y_B \vert \rangle$.

Based on the above findings we now explore the case with a more general baryon-violating coupling where $\lambda_0 \neq 0$ and $\lambda_2 \neq 0$ in \eqref{interaction general}. In terms of the polar variables, we have the generalized potential as
\begin{align}\label{potential_polar_general}
V(R , \theta) = \frac{1}{2}m^2 R^2 + 2R^4 \left[\lambda_0 \cos(4\theta) + \lambda_1 \cos(2\theta) +\lambda_2\right].
\end{align}
One can check that the equation of motion for the angular mode becomes
\begin{align}
\ddot{\theta} + \left(3H + 2 \frac{\dot{R}}{R}\right) \dot{\theta} = -2R^2\left[2\lambda_0 \sin(4\theta) +\lambda_1\sin(2\theta) \right].
\end{align}
Following the same procedure as have done for the special case with $\lambda_0 =0$, $\lambda_1 =0.03$, we arrive at the main conclusion of this work as
\begin{align}\label{Y_B general}
\langle \vert Y_B \vert \rangle \sim \vert Y_B(R_0, \theta_{\rm max}) \vert \sim  \frac{\lambda_{\rm max} R_0^2}{m} \frac{R(t_R)^2}{s(t_R)} \sim  \frac{\lambda_{\rm max} R_0^2}{m} \frac{R_0^2 H^2(t_R)}{m^2 s(t_R)},
\end{align}
where $R_0$ is the VEV of $R$ during inflation and $R(t_R)$ is the amplitude of the radial mode at the beginning of radiation domination. The parameter $\lambda_{\rm max} \equiv \sum_m \tilde{\lambda}_m(\theta_{\rm max})$ means the maximal value of the combination of $\tilde{\lambda}_0 = 4\lambda_0 \sin(4\theta)$ and $\tilde{\lambda}_1 = 2 \lambda_1\sin(2\theta)$ obtained at $\theta = \theta_{\rm max}$. The last estimation in \eqref{Y_B general} applies to the case of reheating led by coherent inflaton oscillation.


\begin{figure}
	\begin{center}
		\includegraphics[width=9.2cm]{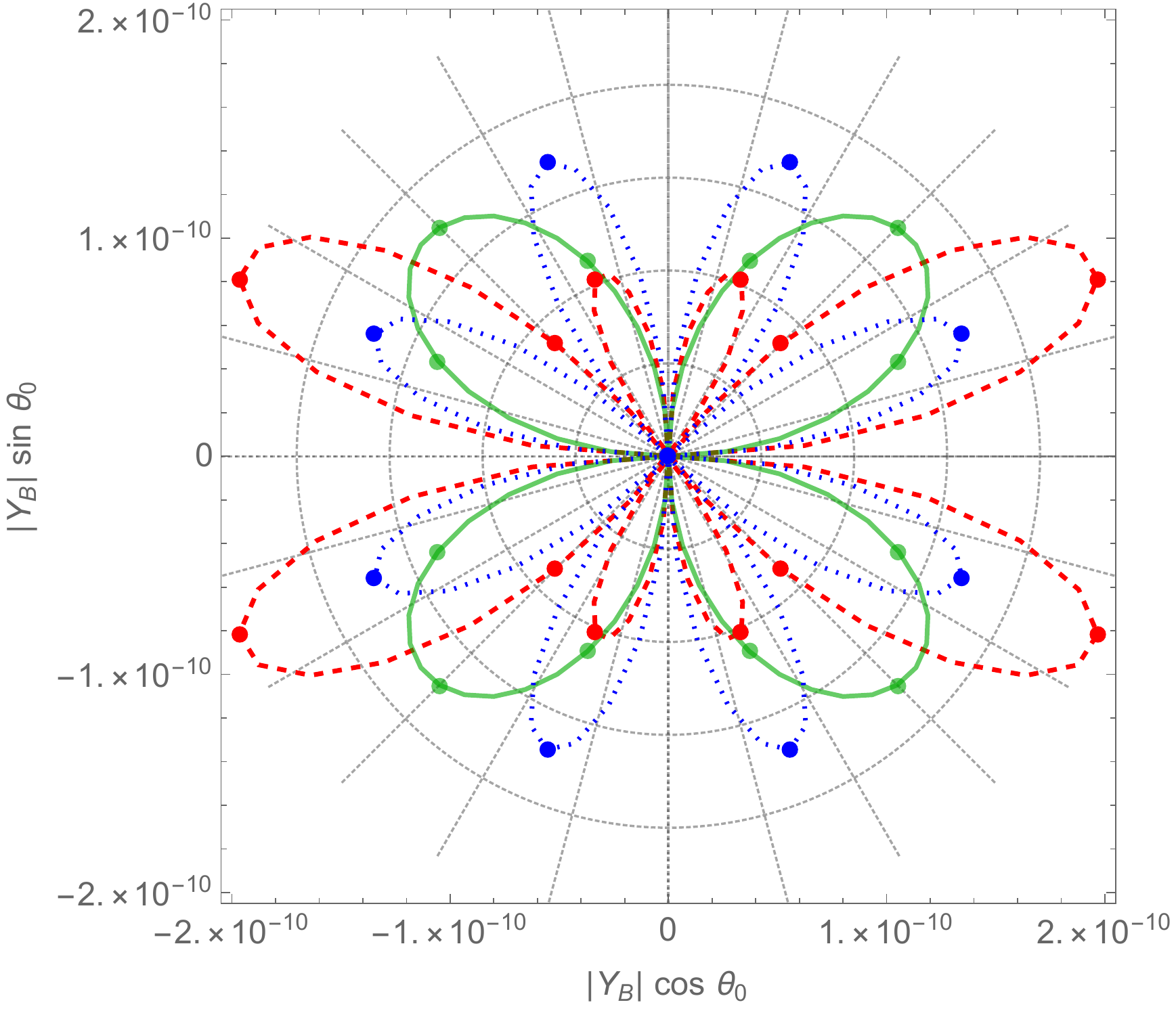}
	\end{center}
	\caption{A scan of the final baryon asymmetry $\vert Y_B\vert$ as a radial function of $\theta_0 \in [ 0, 2\pi ]$ at the fixed value $R_0 = \sqrt{3H_\ast^4/(4\pi^2 m^2)} = r_0H_\ast$, where $Y_B = Y_B(R_0, \theta_0)$ is the numerical result at reheating complete with respect to initial conditions $\{R_0, \theta_0\}$.
		We use $m = 0.5 H_\ast$. 
		The coupling parameters are chozen as $\lambda_0 = \lambda_1 = 0.03$ (red-dashed line), $\lambda_0 = 0$, $\lambda_1 = 0.03$ (green-solid line), and $\lambda_0 = 0.03$, $\lambda_1 = 0$ (blue-dotted line). The dots correspond to $\theta_0 = n\pi/8$. \label{fig:angular_scan_all}}
\end{figure}

In Figure~\ref{fig:angular_scan_all}, we show the scan of angular initial conditions $\theta_0$ for the final baryon asymmetry $\vert Y_B\vert$ with the choice of parameters given by (i) $\lambda_0 = \lambda_1 = 0.03$ as the red-dashed line, (ii) $\lambda_0 =0$, $\lambda_1 = 0.03$ as the green-solid line, and (iii) $\lambda_0 = 0.03$, $\lambda_1 = 0$ as the blue-dotted line. $\vert Y_B\vert$ is the radial distance from the origin as a function of $\theta_0$ at $ r_0 = \sqrt{3H_\ast^2/(4\pi^2 m^2)}$. $\theta_0$ is measured in radians, counterclockwise from the positive $x$ axis. The maximal asymmetry appears around $\theta_0 = n \pi /8$ for the $\tilde{\lambda}_0$ term with $n = 1, 3, \cdots 13, 15$ and for the $\tilde{\lambda}_1$ term with $n = 2, 6, 10, 14$. However at $n = 3, 5, 11, 13$ the sign of $\tilde{\lambda}_0$ is opposite to $\tilde{\lambda}_1$ so that the maximal baryon asymmetry $\vert Y_B \vert$ for the case (i) is generated near $n = 1, 7, 9 ,15$.

Before ending this section, we remark that the generation of baryon asymmetry in this scenario is more like a non-reversible conversion from the source term, the right-hand-side of \eqref{eom: polar_theta}, with spontaneous CP violating initial conditions. 
The dynamics of field oscillation itself does not create the baryon asymmetry if the initial conditions preserves CP. This is the main difference from the spontaneous baryogenesis scenario \cite{Cohen:1987vi} as there is no CPT violating couplings at work.
In the case with a smaller scalar mass $m < H_\ast$, the final baryon asymmetry can have a larger value 
and the enhancement is more efficient in the small mass limit where $m/H_\ast \rightarrow 0$. 

We also remark that the scenario \eqref{AD_Lagrangian} with the $U(1)$ violating interactions \eqref{interaction general} up to quartic order provides a minimal realization of baryogenesis from stochastic initial conditions. This dynamics can be extended to supersymmetric scenarios, which have been the more mainstream implementation of the Affleck-Dine mechanism. In this context, a Hubble-induced tachyonic mass term is typically assumed in order for the scalar field to acquire a large VEV during inflation. However, the Hubble-induced contribution to $m^2$ may be either positive or negative; in fact it is always positive if the Kahler terms are minimal~\cite{Dine:1995kz}. In the case of a positive Hubble-induced contribution to $m^2$, our work implies that the stochasticity may still give rise to successful baryogenesis due to the large value of the inflationary Hubble parameter $H_\ast \sim 10^{13-14}$ GeV. 

\section{Isocurvature constraints}\label{Sec.isocurvature}
The baryon number carried by the charged scalar may convert to ordinary matter via the direct decay of $\phi$ into Standard Model particles, which is familiar in the Affleck-Dine baryogenesis. However, the density fluctuations of baryons generated in this class of scenarios are isocuvature perturbations for the background radiation as the charged scalar is not the inflaton field.  
Observations of the cosmic microwave background have put a constraint on the fraction of the uncorrelated isocuvature perturbations defined as
\begin{align}
\beta_{\rm iso} = \frac{\mathcal{P}_{\rm iso}(k_\ast)}{\mathcal{P}_{\rm iso} (k_\ast)+ \mathcal{P}_\zeta(k_\ast)},
\end{align}
where $\mathcal{P}_\zeta \approx 2.2 \times 10^{-9}$ is the power spectrum of the adiabatic fluctuations, $\mathcal{P}_{\rm iso}$ is the power spectrum of the isocurvature species and $k_\ast$ is the pivot scale of measurements. For $k_\ast = 0.002 \text{Mpc}^{-1}$ the observational constraint shows $\beta_{\rm iso} < 0.021$ \cite{Akrami:2018odb}. This may translate to a constraint to the baryonic fluctuation \cite{Kusenko:2017kdr,Inomata:2018htm} as
\begin{align}
\left\vert \frac{\delta Y_B}{Y_B}\right\vert < \frac{\Omega_{\rm DM}}{\Omega_b} (\beta_{\rm iso} \mathcal{P}_\zeta)^{1/2},
\end{align} 
where for $k_\ast = 0.002 \text{Mpc}^{-1}$ one finds $\vert \delta Y_B/Y_B\vert < 3.4 \times 10^{-5}$. In the limit of $\Omega_b \rightarrow 0$, there is no constraint on the baryon density perturbations.

In the present scenario, the asymmetry $Y_B = n_B /s = R^2\dot{\theta} /s$ where $s$ is given by the adiabatic background temperature. The coherent motion of $\dot{\theta}$ is generated only in the post-inflationary epoch and its value at the end of reheating depends on the initial VEV $\theta_0$. However, $\theta$ is a massless field and thus the VEV $\theta_0$ in each local patch of the Universe is stochastically chosen at the time of horizon crossing during inflation. Note that the quantum fluctuation of a massless field is a constant on superhorizon scales so that the fluctuation of $\dot{\theta}$ is suppressed during inflation. As a result, the perturbation $\delta Y_B$ in a local Universe (which reenters the horizon in the post-inflationary epoch) is mainly sourced by the small fluctuations in the initial VEV $R_0 \equiv \sqrt{\langle R^2\rangle}$. 
\footnote{We focus on a local Universe with a set of initial conditions $\{R_0,\theta_0\}$ that gives the correct baryon asymmetry $Y_B(R_0,\theta_0) = \langle\vert Y_B\vert \rangle$.}
\footnote{A charged scalar condensate may develop inhomogeneity (on smaller physical scales) during the post-inflationary evolution with sufficiently large angular velocity (see, for example \cite{Kusenko:1997si,Dine:2003ax}).}
Namely, the baryonic density perturbation is led by
\begin{align}
\frac{\delta Y_B (x)}{Y_B} = \frac{\delta R^2(x)}{R_0^2} = \frac{R^2(x) - \langle R^2(x) \rangle}{\langle R^2(x) \rangle}.
\end{align}
By virtue of the Fourier expansion for the quadratic perturbation, 
\begin{align}\label{R2 Fourier_expansion}
\delta R^2(x) = \int\frac{d^3 k}{(2\pi)^3} e^{i \vec{k}\cdot\vec{x}} \delta R_{\vec{k}}^2 
= \int \frac{d^3 k_1}{(2\pi)^3}\frac{d^3 k_2}{(2\pi)^3} e^{i (\vec{k}_1+\vec{k}_2)\cdot\vec{x}} 
\left[ R_{\vec{k}_1}R_{\vec{k}_2} - \left\langle R_{\vec{k}_1}R_{\vec{k}_2} \right\rangle \right],
\end{align}
where $\vec{k} = \vec{k}_1 + \vec{k}_2$, we can compute the power spectrum according to
\begin{align}
\left\langle  \delta R_{\vec{k}}^2  \delta R_{\vec{p}}^2\right\rangle &= (2\pi)^3 \delta^3(\vec{k} + \vec{p}) \frac{2\pi^2}{k^3} \mathcal{P}_{\delta R^2}(k) \\
		&= \int d^3x d^3 y e^{-i \vec{k}\cdot \vec{x}} e^{-i \vec{p}\cdot \vec{y}} \left\langle \delta R^2(x) \delta R^2(y)\right\rangle.
\end{align}
Using the Fourier expansion \eqref{R2 Fourier_expansion} and with the help of Wick's theorem, we can express the 4-point correlation function in terms of 2-point correlation functions. A detail derivation can be found, for example, in the Appendix B of \cite{Kusenko:2017kdr}.
After integrating out the spatial coordinates $\vec{x}$ and $\vec{y}$ one obtain the expression in terms of the power spectrum $\mathcal{P}_R(k)$ of the radial mode of the form
\begin{align}\label{R2 spectrum_1}
\left\langle  \delta R_{\vec{k}}^2  \delta R_{\vec{p}}^2\right\rangle  
= 2\delta^3(\vec{k} + \vec{p}) \int d^3k_1  \frac{2\pi^2}{k_1^3} \frac{2\pi^2}{q^3} \mathcal{P}_R(k_1) \mathcal{P}_R(q),
\end{align} 
where $q \equiv \vert \vec{q}\vert$ and $\vec{q} = \vec{k} - \vec{k}_1$.

The information of the power spectrum $\mathcal{P}_R(k)$ resides in the two-point spatial correlation function $G(\vert \vec{x}_1 - \vec{x}_2\vert) = \langle R(\vec{x}_1, t) R(\vec{x}_2, t) \rangle$ at a given time $t$. In general, one constructs the two-point correlation function based on the two-point PDF whose initial condition is fixed by the one-point PDF of $R$ corresponding to the equilibrium state \cite{Starobinsky:1994bd}. 
Noting that the angular dependence led by the baryon violating terms is small, we take the one-point PDF $\rho[R, \theta] \approx \rho[R]$. As a result, the temporal correlation function of $R$ can be readily computed as in the massive non-interacting case in \cite{Starobinsky:1994bd}, which reads 
\begin{align}\label{autocorrelation}
G(\vert t_1 - t_2 \vert) = \left\langle R(\vec{x}, t_1)R(\vec{x}, t_2) \right\rangle 
= \left\langle  R^2\right\rangle e^{-\frac{m^2}{3H_\ast}\vert t_1 - t_2 \vert},
\end{align} 
where $\left\langle  R^2\right\rangle = R_0^2$ is the VEV obtained in the previous section.

In a de Sitter invariant background the temporal correlation function can translate into a spatial correlation function at $t$ via the relation $a(t)H_\ast x = \exp(H_\ast\vert t_1 - t_2 \vert/2) $ \cite{Starobinsky:1994bd,Kunimitsu:2012xx}, where $x \equiv \vert \vec{x}_1 - \vec{x}_2\vert$. We are interested in the spatial correlations over superhorizon scales, namely $x > 1/H_{\rm end}$ at the end of inflation $t = t _{\rm end}$, where $H_{\rm end} = a(t_{\rm end}) H_\ast$. This gives the spatial correlation function as
\begin{align}
G(x) = \int \frac{d^3 k}{(2\pi)^3} P_R(k) e^{-i \vec{k} \cdot \vec{x}}
 = \left\langle  R^2\right\rangle \left(H_{\rm end} x\right)^{-\frac{2m^2}{3H_\ast^2}},
\end{align}
where $P_R(k) = \frac{2\pi^2}{k^3}\mathcal{P}_R(k)$.
Adopting the power-law parametrization $P_R(k) = A_R k^{n}$, the correlation function takes the form of
\begin{align}
G(x) = \frac{A_R}{2\pi^2} x^{-n-3} \Gamma(n + 2) \sin\left[\frac{\pi}{2}(n +2)\right],
\end{align}
which implies $n = -3 + M_\ast$ with $M_\ast \equiv 2m^2/(3H_\ast^2)$ and $A_R = 2\pi^2 c_n \langle  R^2 \rangle H_{\rm end}^{- M_\ast} $ where we define $c_n^{-1} =  \Gamma(n + 2) \sin [(n +2)\pi/2] $. Note that $c_n \rightarrow M_\ast$ as $M_\ast \rightarrow 0$.

The power spectrum of $R$ at the end of inflation is therefore given by
\begin{align}
\mathcal{P}_R(k) = c_n \left\langle  R^2\right\rangle \left(\frac{k}{H_{\rm end} }\right)^{-M_\ast}.
\end{align}
Inserting this result back into \eqref{R2 spectrum_1} with $\langle R^2 \rangle = R_0^2$ leads to
\begin{align}\label{R2 spectrum_2}
\left\langle  \delta R_{\vec{k}}^2  \delta R_{\vec{p}}^2\right\rangle  
= 2c_n^2 R_0^4 \delta^3(\vec{k} + \vec{p}) \int d^3k_1  \frac{2\pi^2}{k_1^3} \frac{2\pi^2}{q^3} \left(\frac{k_1 q}{H_{\rm end}} \right)^{M_\ast},
\end{align} 
which indicates the power spectrum
\begin{align}
\mathcal{P}_{\delta R^2} (k) = \frac{k^3}{2\pi} c_n^2 R_0^4 \int dk_1^3 \frac{1}{k_1^3 q^3} \left(\frac{k_1 q}{H_{\rm end}} \right)^{M_\ast}.
\end{align}
To work out this integration it is more convenient to use the dimensionless variables $u = k_1/k$ and $v = q / k$. The relation $q^2 = k^2 + k_1^2 -2k k_2 \cos \beta$ transforms to $v = (1 + u^2 -2u \cos\beta)^{1/2}$, where $\beta$ is the angle between $\vec{k}$ and $\vec{k}_1$.
The volume element $d^3k_1 = k^3 d^3u = 2\pi k^3 u^2 du (-d\cos\beta)$. After integrating out $d\cos\beta$, we find
\begin{align}\label{R2 spectrum_final}
\mathcal{P}_{\delta R^2} (k) =  c_n^2(M_\ast) R_0^4 \left(\frac{k }{H_{\rm end}} \right)^{2M_\ast}\mathcal{ F}(M_\ast),
\end{align}
with the numerical function
\begin{align}
\mathcal{ F}(M_\ast) = \int_{1}^{\infty}  u^{M_\ast - 1} \frac{(u - 1)^{M_\ast -1} - (u+1)^{M_\ast - 1}}{u(1- M_\ast)} \,du.
\end{align}
We note that modes with $k_1 < k$ have become superhorizon background before the $k$-mode crosses the horizon so that they do not contribute to the perturbation of $\delta R^2_{\vec{k}}$. Therefore the lower limit of the integration of $\mathcal{ F}(M_\ast)$ is $u= k_1/k = 1$.

\begin{figure}
	\begin{center}
		\includegraphics[width=7 cm]{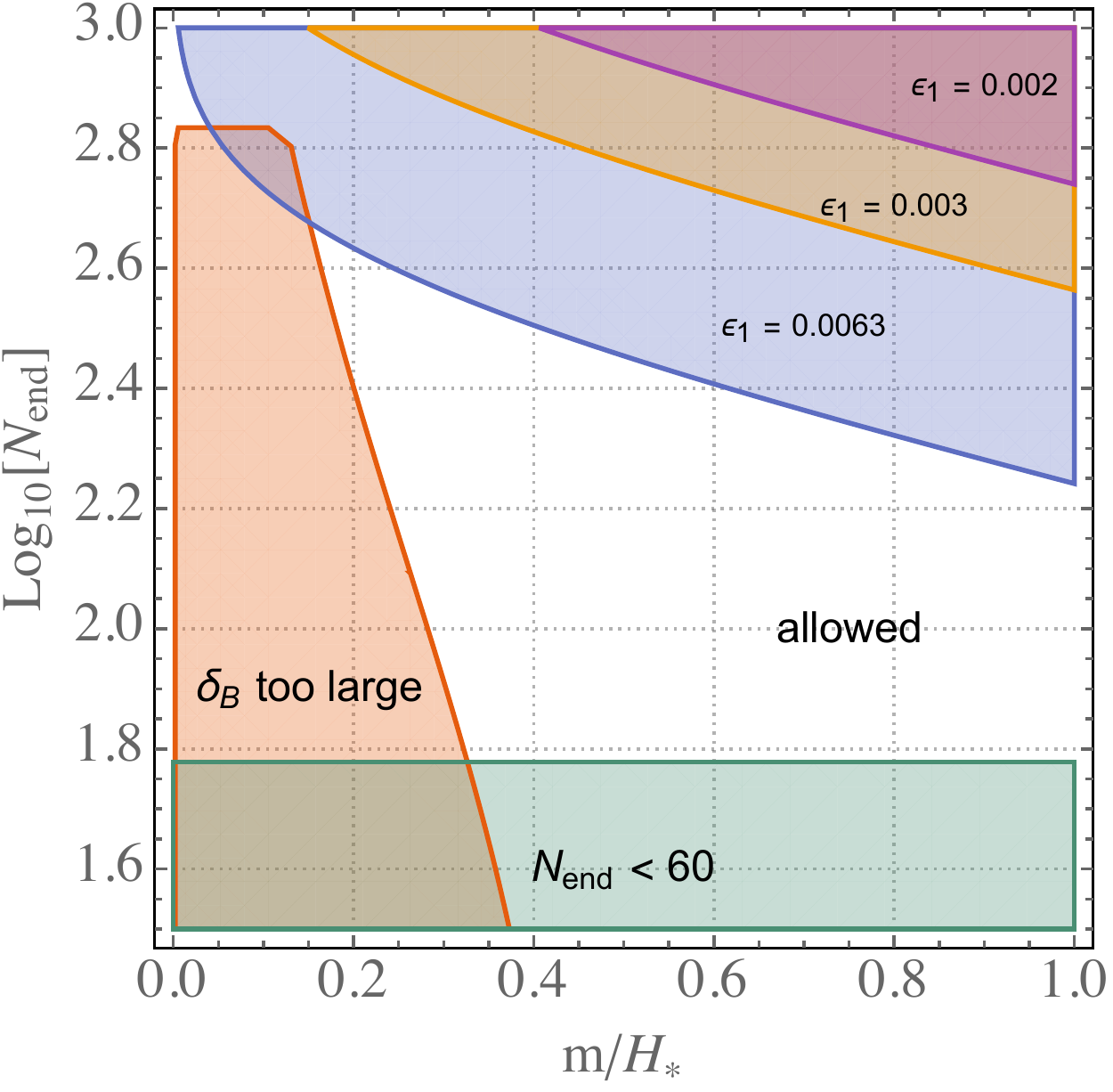}
	\end{center}
	\caption{The parameter scan of the baryon density perturbation $\delta_B(m, N_{\rm end})$ with respect to the scalar mass $m$ and the number of $e$-folds to the end of inflation $N_{\rm end}$, where the red-shaded region is excluded by the isocurvature constraint $\vert \delta_B \vert < 3.4 \times 10^{-5}$.
	 The allowed parameter space constrained by non-oscillation condition $H_{\rm end} > m/3$ in slow-roll inflation is shown with various choices of the parameter $\epsilon_1 \equiv - \dot{H}/H^2$, up to the current upper bound $\epsilon_1 = 0.0063$~\cite{Akrami:2018odb}.
		\label{fig:isocurvature}}
\end{figure}

Finally we obtain the baryonic fluctuation independent of the baryon violating couplings as
\begin{align}
\delta_B = \frac{\delta Y_B}{Y_B} \approx \frac{\mathcal{P}_{\delta R^2}^{1/2}}{R_0^2} 
= c_n(M_\ast) \left(\frac{k }{H_{\rm end}} \right)^{M_\ast} \mathcal{ F}(M_\ast)^{1/2}.
\end{align}
In the reheating scenario \eqref{reheating background}, the ratio of the scale factor from inflation end to the beginning of radiation domination is $a_R/a_{\rm end} = (\Lambda_I / T_R )^{4/3} $, where $T_R = T(t_R)$ is the temperature around reheating completed. Assuming that the entropy of the Universe is conserved after reheating, we can connect the scale factor at present $a_0$ to $a_R \equiv a(t_R)$ through $a_0^3 s(T_0) = a_R^3 s(T_R)$, where $T_0 = 2.73 K \approx 2.35 \times 10^{-13}$ GeV. We use the Standard Model value $g_{\ast 0} \equiv g_{\ast}(T_0) = 43/11$ for $s(T_0) = 2\pi^2 g_{\ast 0}T_0^3/45$. Setting $a_0 = 1$, the perturbation mode $k$ in the unit of $H_{\rm end}$ is now given by
\begin{align}\label{k_in_H_end}
\frac{k}{H_{\rm end}} = 2\pi e^{-N_{\rm end}}  \left(\frac{T_R}{\Lambda_I}\right)^{4/3} \left(\frac{g_{\ast 0}}{g_\ast}\right)^{1/3}\frac{T_0}{T_R},
\end{align}
where $N_{\rm end}$ means the number of $e$-folds required from $k$-mode exits the horizon to the end of inflation. As a result, the isocurvature constraint $\vert \delta_B \vert < 3.4 \times 10^{-5}$ restricts the parameter space for $N_{\rm end}$ and $M_\ast$ (namely $m$) as shown in Figure~\ref{fig:isocurvature}. 
\footnote{The parameter space for $m/H_\ast < 0.0021$ in Figure~\ref{fig:isocurvature} is in fact compatible with the isocurvature constraint $\vert \delta_B \vert < 3.4 \times 10^{-5}$ due to the vanishing of the coefficient $c_n$ in the small mass limit. In this limit the spatial correlation length $x_c$ defined as $G(x_c) = \langle R^2\rangle/2$ \cite{Kunimitsu:2012xx, Starobinsky:1994bd} becomes exponentially larger than the horizon scale, see also the discussion in \cite{Dine:1995kz,Hook:2015foa}.}
 One can see that the baryonic fluctuation is always compatible with observations for a mass $m \geq 0.4 H_\ast$ with a minimal duration of inflation $N_{\rm end} > 60$. For $m < 0.1 H_\ast$ one would need $N_{\rm end} > 10^{2.8}$ to pass the isocurvature constraint.

So far we have treated the Hubble parameter $H = H_\ast$ as a constant throughout inflation. However, in realistic case $H$ may decrease slightly during inflation. If $H_{\rm end} < m/3$, then the scalar field would start oscillating before the end of inflation and the generated baryon asymmetry would be diluted. For single-field inflationary models, the first slow-roll parameter $\epsilon_1 = -\dot{H}/H^2$ is constrained to be  $0\leqslant \epsilon_1 < 0.0063$~\cite{Akrami:2018odb}. Requiring that $H_{\rm end} = H_\ast e^{-\epsilon_1 N_{\rm end}} > m/3 > 0.4 H_\ast/3$ implies $N_{\rm end} < 2/\epsilon_1$, which gives the most stringent constraint $N_{\rm end} < 320$ for the largest allowed value of $\epsilon_1$. Thus, except for models that require $N_{\rm end} \gg 10^2$, the observed baryon asymmetry can be obtained in generic single-field slow-roll inflation for $m > 0.4 H_\ast$. 
Note that for slow-roll inflation the factor in \eqref{k_in_H_end} is marginally modified as $e^{-N_{\rm end }} \rightarrow e^{- (1-\epsilon_1) N_{\rm end}}$.


\section{Summary}\label{Sec.summary}
We have revisited baryogenesis, in the framework of the Affleck-Dine mechanism, via a charged scalar field, $\phi = Re^{i\theta}/\sqrt{2}$, whose initial VEV during inflation is determined by an extended stochastic method with multi-field configuration and broken continuous symmetry. The probability distribution of VEVs in general depends on both the radial direction $R$ and the phase direction $\theta$, where the latter carries information of the spontaneous CP violation. 
Due to the smallness of the baryon-number-violating couplings, $\lambda_m$, the $\theta$ distribution investigated in this work is nearly homogeneous and thus the angular VEV $\theta_0 $ behaves as a pure stochastic variable during inflation. 

We have shown that the final baryon asymmetry $Y_B$ in a local Universe significantly depends on the initial value of $\theta_0$, given that CP violation is one of the three necessary conditions to obtain a non-zero baryon number. On the other hand, the global expectation value, $\langle\vert Y_B\vert\rangle$, averaging over all values of $R$ and $\theta$ is dominated by the local contribution $Y_B = Y_B(R_0, \theta_{\rm max})$ from the initial VEVs with maximal CP violation $\theta_0 = \theta_{\rm max}$. Note that $R_0 = \langle R^2\rangle^{1/2}$ is simply the VEV of a massive scalar (without $U(1)_\phi$ symmetry breaking). We also remark that $\theta_{\rm max}$ depends on the choices of the coupling constants $\lambda_m$, and that the global value $\langle\vert Y_B\vert\rangle$ should be considered as the true physical observable of the presented scenario with stochastic initial conditions.

Finally, we emphasize that $\langle\vert Y_B\vert\rangle$ is a pure number independent of the correlation length of the radial VEV $R_0$ or any other cosmological scale, whereas the local asymmetry $Y_B(R_0,\theta_0)$ based on a set of particular initial VEVs $\{R_0,\theta_0\}$ is indeed characterized by the length scale of $R_0$ \cite{Dolgov:1991fr,Hook:2015foa}. In other words, different patches of the Universe with different initial values of $R_0$ will reach to the same conclusion as $\langle\vert Y_B\vert\rangle$. 
We have considered the local baryonic fluctuation as a fully uncorrelated isocurvature perturbation to the adiabatic fluctuations of the CMB radiations and found that a scalar mass $m \sim\mathcal{O}(H_\ast)$ can be compatible with observational constraints.

\acknowledgments
We thank Alexander Kusenko for helpful discussions.
This work was supported by the ANR ACHN 2015 grant (``TheIntricateDark'' project), and by the NWO Vidi grant ``Self-interacting asymmetric dark matter.''

\appendix
\section{Single-field stochastic inflation}\label{Appendix A}


In the standard approach of stochastic inflation \cite{Starobinsky:1994bd}, a quantum field in the de Sitter background is decomposed into a long-wavelength part and a short-wavelength counterpart. The short-wavelength part is a collection of quantum modes well inside the Hubble horizon and it serves as a noise term for the stochastic motion of the long-wavelength component on superhorizon scales. The probability distribution of the long-wavelength component coarse-grained around the Hubble horizon can be resolved from the Fokker-Planck equation and the distribution function is in general time-dependent.
In this section we focus on the stationary solution of the probability distribution function in a field potential of (meta)stable shape.

\subsection{Basic formalism}
We review in this subsection the basic equations for computing the field condensation of a single scalar in de-Sitter space \cite{Starobinsky:1994bd}.
In the picture of stochastic inflation the superhorizon fluctuations of a scalar $\phi$ comprise as a coherent (auxiliary) field, denoted by $\phi_0$. The one-point probability distribution function (PDF) of $\phi_0$, namely $\rho_1[\phi_0(\mathbf{x}, t)]$, satisfies the Fokker-Planck equation \cite{Starobinsky:1994bd}
\begin{equation}\label{FP eq1}
\frac{\partial\rho_1[\phi_0]}{\partial t} =
\frac{1}{3H} \frac{\partial}{\partial \phi_0} 
\left( \rho_1[\phi_0]\, \frac{\partial V_{\rm eff}}{\partial \phi_0}  \right) +
\frac{H^3}{8\pi^2}
\frac{\partial^2\rho_1[\phi_0]}{\partial \phi_0^2} .
\end{equation}
One can construct the two-point or higher-point PDFs and correlation functions of the scalar based on the one-point PDF.
In terms of the dimensionless variables $\tau\equiv Ht$, $x \equiv \phi_0/H$, $U = V/H^4$, we rewrite the Fokker-Planck equation as
\begin{align}
\frac{\partial\rho_1[x]}{\partial \tau} = \frac{1}{3} \frac{\partial}{\partial x} 
\left( \rho_1[x]\, \frac{\partial U(x)}{\partial x}  \right) +
\frac{1}{8\pi^2} \frac{\partial^2\rho_1[x]}{\partial x^2}\equiv \mathbb{L}\, \rho_1[x].
\end{align}
The right-hand side can be written as the divergence of a current $J(x, \tau)$ \cite{Risken:1984}, such that
\begin{align}\label{FP eq2}
\mathbb{L}\, \rho_1[x] = \frac{1}{8\pi^2}\frac{\partial}{\partial x}
\left[e^{-8\pi^2U/3}\frac{\partial}{\partial x}\left(e^{8\pi^2 U/3}\rho_1[x]\right)\right]\equiv -\frac{\partial}{\partial x} J( x, \tau),
\end{align}
where $J$ is the probability current of the form
\begin{equation}\label{probability current}
J(x, \tau)=- \frac{1}{8\pi^2}e^{-8\pi^2 U/3}\frac{\partial}{\partial x}\left(e^{8\pi^2 U/3}\rho_1[x]\right).
\end{equation}
Then the Fokker-Planck equation takes the form of $\partial_\tau \rho_1 + \vec{\nabla} J =0$.
The general solution of the one-point PDF $\rho_1$ can be written as
\begin{equation}\label{PDF generic solution}
\rho_1(x, \tau)=\sum\limits_{n=0}^{\infty}
\alpha_n\Psi_n(x)e^{-\Lambda_n(\tau-\tau_0)},
\end{equation}
where $\Psi_n(x)$ denotes a complete set of eigenfunctions of the Fokker-Planck operator $\mathbb{L}$ with the eigenvalues $- \Lambda_n$.
The definition of \eqref{PDF generic solution} implies that $\Lambda_n$ should be non-negative, or otherwise the $n$-th state solution is unstable.
 Taking \eqref{PDF generic solution} into \eqref{FP eq2}, we obtain the eigenfunction equations
\begin{equation}
\mathbb{L}\,\Psi_n(x)=-\Lambda_n\Psi_n(x),
\end{equation}
with suitable boundary conditions depending on the shape of the potential $U(x)$. One can see that $\Lambda_n$ mimics the energy level of $\Psi_n(x)$ in the Schr\"{o}dinger-type equation with a discrete spectrum due to the boundary conditions of the field potential. 

Note that $\mathbb{L}$ defined in \eqref{FP eq2} is not Hermitian, yet a Hermitian operator can be constructed out of $\mathbb{L}$ as, for example, $\mathbb{L}_H \equiv e^{Q/2}\mathbb{L}e^{-Q/2}$ with $Q = 8\pi^2 U/3$ \cite{Risken:1984}. For a set of solutions $\Psi_n$ satisfying the same boundary condition $J(x_b) = 0$ at a boundary point $x = x_b$, it is easy to show the self-adjoint of the operator $\Psi_m \mathbb{L}_H \Psi_n = \Psi_n \mathbb{L}_H \Psi_m$ by using integration by part for two times. One can also consider another complete set of eigenfunctions $\Phi_n \equiv e^{Q/2}\Psi_n$ with respect to $\mathbb{L}_H$, which share the same eigenvalues with $\mathbb{L}$ as $\mathbb{L}_H \Phi_n(x) = -\Lambda_n\Phi_n(x)$.
Therefore the orthonormal relation for $\Phi_n$ give rise to the relation among $\Psi_n$ as 
\begin{equation}
\label{orthonormal relation}
\int^{x_1}_{x_2} \Phi_m(x)\Phi_n(x)\, dx = \int^{x_1}_{x_2} e^{Q}\Psi_m(x)\Psi_n(x)\, dx=\delta_{mn}.
\end{equation}
Note that $\Phi_n$ is the eigenfunction used in \cite{Starobinsky:1994bd} and \cite{Adshead:2020ijf}.

It is sometimes convenient to rewrite the Hermitian operator in the form of
\begin{align}\label{Hermitian_L_ladder}
\mathbb{L}_H = \frac{1}{8\pi^2}e^{Q/2}\frac{\partial}{\partial x} \left[e^{-Q} \frac{\partial}{\partial x}  e^{Q/2}\right] 
\equiv - \mathbb{A}^\dagger\mathbb{A},
\end{align} 
where $\mathbb{A}$ and its hermitian conjugate $\mathbb{A}^\dagger$ behave as the ladder operator for the energy level of the system, and they are defined by
\begin{align}\label{ladder_operator}
\mathbb{A} = \sqrt{\frac{1}{8\pi^2}} e^{-Q/2} \frac{\partial}{\partial x} e^{Q/2}, \qquad 
\mathbb{A}^\dagger = -\sqrt{\frac{1}{8\pi^2}} e^{Q/2} \frac{\partial}{\partial x} e^{-Q/2},
\end{align}
respectively. The ladder operators satisfy the commutation relation $[\mathbb{A}, \mathbb{A}^\dagger] = \frac{1}{8\pi^2}\partial_x^2 Q$. For a quadratic potential $\partial_x^2 Q$ is a non-zero constant and one can obtain higher-level solutions from the ground state function $\Phi_0$ by virtue of the ladder operators.

\subsection{Equilibrium states}
If there is enough time for the PDF to reach the stationary condition $\partial \rho_1/\partial t = 0$, then $J$ becomes a constant in $x$. This implies $J = 0$ everywhere if the probability current vanishes at the boundaries of the potential. The solution of $J = 0$ corresponds to an equilibrium state obtained from \eqref{probability current} as
\begin{align}
\rho^{\rm eq}_1 = e^{-Q}/\Sigma,
\end{align}
where $\Sigma = \int_{x_{b2}}^{x_{b1}} e^{-Q} dx$ is a constant that ensures the normalization condition $\int \rho^{\rm eq}_1 dx = 1$ in the given region $x \in [ x_{b1},x_{b2} ]$.

For a bounded potential where the probability current vanishes at $x\rightarrow \pm \infty$, the equilibrium state is nothing but the ground state solution with $\Lambda_0 = 0$.
Thus we decompose \eqref{PDF generic solution} as
\begin{align}
\rho_1 (x,\tau) = \rho^{\rm eq}_1(x) + \sum_{n = 1}^{\infty} \alpha_n \Psi_n(x)e^{-\Lambda_n (\tau -\tau_0)},
\end{align}
where $\rho^{\rm eq}_1(x) = \alpha_0\Psi_0(x)$ is a constant in $\tau$. For a symmetric potential vanishes at the origin, or namely $U(0) = 0$, the ground state eigenfunction takes the form $\Psi_0(x) = e^{-8\pi^2U/3}\Psi_0(0)$. The orthonormal condition \eqref{orthonormal relation} then implies $\Psi_0(0) = 1/\sqrt{\Sigma}$. Since $\rho^{\rm eq}_1(x)$ is the asymptotic solution at $\tau \rightarrow \infty$, we find $\alpha_0 = 1/\sqrt{\Sigma}$ by virtue of the normalization condition $\int_{-\infty}^{\infty} \rho^{\rm eq}_1 dx = 1$. 

Let us consider an example given by the effective potential for $\phi_{I}$ in \eqref{potential_leading_order}. In terms of the dimensionless variables the potential reads
\begin{align}\label{stable potential}
U(x) = \frac{\lambda}{4} x^4 + \frac{1}{2}M^2 x^2,
\end{align}
where $\lambda$ and $M = m/H_\ast$ are also dimensionless parameters.  
Here the normalization constant can be evaluated as
\begin{align}
\Sigma = \int_{-\infty}^{\infty} e^{-8\pi^2U/3} dx = \sqrt{\frac{M^2}{2\lambda}} \exp\left(\frac{\pi^2 M^4}{3\lambda}\right) 
K_{\frac{1}{4}}\left(\frac{\pi^2 M^4}{3\lambda}\right) ,
\end{align}
where $K_n$ is the modified Bessel function of the second kind.
For the parameter space of our main interest, we have $\pi^2 M^4/(3\lambda) \gg 1$ so that 
\begin{align}
\Sigma \approx \sqrt{\pi} x_0\left(1 - \frac{1}{2} \pi^2\lambda x_0^4\right),
\end{align}
with $x_0^2 = 3/(4\pi^2 M^2)$. 

In the limit of $\tau\rightarrow \infty$, the expectation value approaches
\begin{align}
\left\langle x^2\right\rangle = \int_{-\infty}^{\infty} x^2 \rho_1^{\rm eq} (x) dx \approx \frac{3}{8\pi^2 M^2}\left(1-\frac{9}{8} \frac{\lambda}{\pi^2 M^4}\right).
\end{align}
The result for the massive non-interacting scenario $\langle x^2\rangle = 3/(8\pi^2 M^2)$ \cite{Starobinsky:1994bd} is recovered at the limit of $\lambda = 0$.

\subsection{Quasi-equilibrium states}
In the case the lowest eigenvalue $\bar{\Lambda}_0 \ll 1$ is a small but non-zero value, the late-time solution led by \eqref{PDF generic solution} at $\tau \gg 1/ \bar{\Lambda}_0$ is quasi-stationary. This is the case for a meta-stable potential $\bar{U}(x)$ with a sufficiently large potential barrier $\bar{U}(x_\ast) \gg 1$, where $x_\ast$ is the field value at the local maximum of the potential.  The escape rate from inside the potential is proportional to $J(x_\ast)$ and can be approximated by the lowest non-vanishing eigenvalue $\bar{\Lambda}_0$ \cite{Risken:1984}. Here we use $\bar{\Lambda}_n$ to denote the eigenvalues for the general solution of the meta-stable potential $\bar{U}(x)$. As long as the relaxation time scale $\tau_{\rm rex} \equiv \ln 2/\bar{\Lambda}_0$ \cite{Starobinsky:1994bd} is much greater than the duration of inflation, the eigenstate $\bar{\Psi}_0$ corresponds to $\bar{\Lambda}_0$ can be taken as a quasi-equilibrium state. $\bar{\Psi}_0$ is solved according to the eigenfunction equation  
$\mathbb{L}\,\bar{\Psi}_0(x)=-\bar{\Lambda}_0\bar{\Psi}_0(x)$:
\begin{align}
\frac{1}{8\pi^2}\frac{\partial}{\partial x}
\left[e^{-8\pi^2\bar{U}/3}\frac{\partial}{\partial x}\left(e^{8\pi^2 \bar{U}/3}\bar{\Psi}_0(x)\right)\right] = -\bar{\Lambda}_0\bar{\Psi}_0(x).
\end{align}
If $\bar{U}(x)$ is symmetric in $x \leftrightarrow -x$, the derivative $\partial \bar{U}/\partial x$ vanishes at $x = 0$. The generic solution for $\bar{\Psi}_0$ is therefore given by the integration form as
\begin{align}\label{Psi_0 general}
\bar{\Psi}_0(x) = e^{-8\pi^2\bar{U}(x)/3}\left\{\bar{\Psi}_0(0) - 8\pi^2\bar{\Lambda}_0 \int_{0}^{x}d x_2 \left[e^{8\pi^2\bar{U}(x_2)/3}\int_{0}^{x_2}d x_1\bar{\Psi}_0(x_1)\right]\right\}.
\end{align} 
For a very small $\bar{\Lambda}_0$ due to the large barrier $8\pi^2 \bar{U}(x_\ast) \gg 1$, we perform the expansion $\bar{\Lambda}_0 = \bar{\Lambda}_0^{(0)} + \bar{\Lambda}_0^{(1)} +\cdots$ with $\bar{\Lambda}_0^{(0)} = 0$. This implies the zeroth order solution 
\begin{align}
\bar{\Psi}_0^{(0)} (x) = e^{-8\pi^2\bar{U}(x)/3}\bar{\Psi}_0(0).
\end{align}
Inserting the zeroth approximation back into \eqref{Psi_0 general} gives the first order solution as
\begin{align}
\bar{\Psi}_0^{(1)} (x) = e^{-8\pi^2 \bar{U}(x)/3}\bar{\Psi}_0(0) \left[1- \bar{\Lambda}_0^{(1)} f_1(x) \right],
\end{align}
where $f_1(x) = 8\pi^2 \int_{0}^{x}d x_1 (e^{8\pi^2\bar{U}(x_1)/3}\int_{0}^{x_1}d x_2 e^{-8\pi^2\bar{U}(x_2)/3})$.

Let us consider the effective potential \eqref{potential_leading_order} for $\phi_R$ corresponds to the meta-stable shape of the form
\begin{align}\label{U_metastable}
\bar{U}(x) = -\frac{\lambda}{4} x^4 + \frac{1}{2}M^2 x^2,
\end{align}
where the maximum of the potential $\bar{U}_{\rm max} = \bar{U}(x_\ast)$ is given by $x_\ast = \pm \sqrt{M^2/\lambda}$.
The eigenvalue can be fixed by the boundary condition $\bar{\Psi}_0(x_A) = 0$, which indicates $\bar{\Lambda}_0^{(1)}  = 1/f_1(x_A)$.
For the metastable potential \eqref{U_metastable}, the largest possible boundary value $x_A = \phi_A/H$ can be approximated by the critical point at which the classical motion $\Delta\phi_c \approx \vert \dot{\phi}\Delta t \vert = \vert \partial_\phi V/(3H^2)\vert $ starts to dominate the quantum fluctuations $\Delta\phi_q\approx\sqrt{\langle \phi^2\rangle}$. This critical point is lightly larger than $x_\ast$ for $M \gg 1$, and for conservative we adopt $\vert x_A\vert \leq \vert x_\ast \vert = \sqrt{M^2/\lambda}$. For $x_\ast \gg 1$, the largest possible eigenvalue reads
\begin{align}\label{eigenvalue_meta_0}
\bar{\Lambda}_0^{(1)}  =  \frac{1}{f(x_A)} \leq  \sqrt{\frac{2}{9\pi^2}} M^2\exp\left(-\frac{2\pi^2 M^4}{3\lambda}\right).
\end{align}
The existence of a stable state requires $\bar{\Lambda}_0 \Delta\tau \ll 1$ where $\Delta\tau = \tau - \tau_0 \sim \mathcal{O}(60)$ is the duration of inflation. 

The quasi-equilibrium state $\rho_1^{\rm qeq}(x,\tau) = \bar{\alpha}_0\bar{\Psi}_0(x)e^{-\Lambda_0(\tau - \tau_0)} \approx \bar{\alpha}_0\bar{\Psi}_0(x)$ that satisfies the boundary condition $\int_{-x_A}^{x_A}\rho_1^{\rm qeq}(x)dx = 1$ indicates $\bar{\alpha}_0 = \bar{\Psi}_0(0) \approx 1/\sqrt{\Sigma}$. The normalization constant can be computed as
\begin{align}
\Sigma &= \int_{-x_A}^{x_A} e^{-8\pi^2U(x)/3} dx, \\ \nonumber
&\approx \int_{-x_A}^{x_A} e^{-x^2/x_0^2}\left(1+ \frac{2\lambda }{3} \pi^2 x^4\right) dx, \\
& = \sqrt{\pi} x_0\left(1 + \frac{\lambda }{2} \pi^2x_0^4\right),
\end{align}
where $x_0^2 = 3/(4\pi^2 M^2)$. Note that this approximation is valid for the boundary value $(3/(2\pi^2 \lambda))^{1/4} \gg x_A \gg x_0$ and $x_A \ll x_\ast$. 
The expectation value is then
\begin{align}
\left\langle x^2\right\rangle = \int_{-x_A}^{x_A} x^2 \rho_1^{\rm qeq} (x) dx \approx 
\frac{3}{8\pi^2 M^2}\left(1+\frac{9}{8} \frac{\lambda}{\pi^2 M^4}\right).
\end{align}

Finally, we remark that the lowest eigenvalue $\bar{\Lambda}_0$ for the meta-stable potential $\bar{U}(x)$ is in fact the lowest non-zero eigenvalue $\Lambda_1$ for the inverted bistable potential $U(x) = - \bar{U}(x)$. To see this, we denote $\bar{Q} = - Q$ for the meta-stable potential $\bar{U}$ such that the ladder operators \eqref{ladder_operator} transform as $\bar{\mathbb{A}} = - \mathbb{A}^\dagger$ and $\bar{\mathbb{A}}^\dagger = - \mathbb{A}$. 
As a result, the Hermitian operator \eqref{Hermitian_L_ladder} for the meta-stable potential is $\bar{\mathbb{L}}_H = -\bar{\mathbb{A}}^\dagger \bar{\mathbb{A}} = -\mathbb{A} \mathbb{A}^\dagger$.
Putting $\mathbb{A}$ to the left of the eigenvalue equation leads to
\begin{align}
\mathbb{A} \,\mathbb{L}_H \Phi_n = - \mathbb{A}\, \mathbb{A}^\dagger \mathbb{A} \Phi_n 
= \bar{\mathbb{L}}_H \mathbb{A} \Phi_n = -\Lambda_n \mathbb{A} \Phi_n,
\end{align}
which implies that $\mathbb{A} \Phi_n \sim \Phi_{n-1} \sim \bar{\Phi}_m$ is the eigenfunction of $\bar{\mathbb{L}}_H$. One can check that the  eigenvalue given by \eqref{eigenvalue_meta_0} for the meta-stable potential \eqref{U_metastable} agrees with the lowest non-zero value $\Lambda_1$ found in \cite{Starobinsky:1994bd} obtained from the double-wall (bistable) potential.




\begin{thebibliography}{99}
\bibitem{Akrami:2018odb}
Y.~Akrami \textit{et al.} [Planck],
[arXiv:1807.06211 [astro-ph.CO]].

\bibitem{Starobinsky:1986fx}
A.~A.~Starobinsky,
Lect. Notes Phys. \textbf{246}, 107-126 (1986).

\bibitem{Rey:1986zk}
S.~J.~Rey,
Nucl. Phys. B \textbf{284}, 706-728 (1987).

\bibitem{Sasaki:1987gy}
M.~Sasaki, Y.~Nambu and K.~i.~Nakao,
Nucl. Phys. B \textbf{308}, 868-884 (1988).

\bibitem{Nambu:1988je}
Y.~Nambu and M.~Sasaki,
Phys. Lett. B \textbf{219}, 240-246 (1989).

\bibitem{Morikawa:1989xz}
M.~Morikawa,
Phys. Rev. D \textbf{42}, 1027-1034 (1990).

\bibitem{Mollerach:1990zf}
S.~Mollerach, S.~Matarrese, A.~Ortolan and F.~Lucchin,
Phys. Rev. D \textbf{44}, 1670-1679 (1991).


\bibitem{Linde:1991sk}
A.~D.~Linde,
Nucl. Phys. B \textbf{372}, 421-442 (1992)
[arXiv:hep-th/9110037 [hep-th]].

\bibitem{Starobinsky:1994bd} 
A.~A.~Starobinsky and J.~Yokoyama,
Phys.\ Rev.\ D {\bf 50}, 6357 (1994)
[astro-ph/9407016].

\bibitem{Tsamis:2005hd}
N.~C.~Tsamis and R.~P.~Woodard,
Nucl. Phys. B \textbf{724}, 295-328 (2005)
[arXiv:gr-qc/0505115 [gr-qc]].

\bibitem{Finelli:2008zg}
F.~Finelli, G.~Marozzi, A.~A.~Starobinsky, G.~P.~Vacca and G.~Venturi,
Phys. Rev. D \textbf{79}, 044007 (2009)
[arXiv:0808.1786 [hep-th]].

\bibitem{Finelli:2010sh}
F.~Finelli, G.~Marozzi, A.~A.~Starobinsky, G.~P.~Vacca and G.~Venturi,
Phys. Rev. D \textbf{82}, 064020 (2010)
[arXiv:1003.1327 [hep-th]].

\bibitem{Garbrecht:2013coa}
B.~Garbrecht, G.~Rigopoulos and Y.~Zhu,
Phys. Rev. D \textbf{89}, 063506 (2014)
[arXiv:1310.0367 [hep-th]].

\bibitem{Garbrecht:2014dca}
B.~Garbrecht, F.~Gautier, G.~Rigopoulos and Y.~Zhu,
Phys. Rev. D \textbf{91}, 063520 (2015)
[arXiv:1412.4893 [hep-th]].

\bibitem{Onemli:2015pma}
V.~K.~Onemli,
Phys. Rev. D \textbf{91}, 103537 (2015)
[arXiv:1501.05852 [gr-qc]].

\bibitem{Fujita:2013cna}
T.~Fujita, M.~Kawasaki, Y.~Tada and T.~Takesako,
JCAP \textbf{12}, 036 (2013)
[arXiv:1308.4754 [astro-ph.CO]].

\bibitem{Fujita:2014tja}
T.~Fujita, M.~Kawasaki and Y.~Tada,
JCAP \textbf{10}, 030 (2014)
[arXiv:1405.2187 [astro-ph.CO]].



\bibitem{Vennin:2015hra}
V.~Vennin and A.~A.~Starobinsky,
Eur. Phys. J. C \textbf{75}, 413 (2015)
[arXiv:1506.04732 [hep-th]].


\bibitem{Assadullahi:2016gkk}
H.~Assadullahi, H.~Firouzjahi, M.~Noorbala, V.~Vennin and D.~Wands,
JCAP \textbf{06}, 043 (2016)
[arXiv:1604.04502 [hep-th]].

\bibitem{Vennin:2016wnk}
V.~Vennin, H.~Assadullahi, H.~Firouzjahi, M.~Noorbala and D.~Wands,
Phys. Rev. Lett. \textbf{118}, no.3, 031301 (2017)
[arXiv:1604.06017 [astro-ph.CO]].

\bibitem{Kitamoto:2018dek}
H.~Kitamoto,
Phys. Rev. D \textbf{100}, no.2, 025020 (2019)
[arXiv:1811.01830 [hep-th]].

\bibitem{Pinol:2018euk}
L.~Pinol, S.~Renaux-Petel and Y.~Tada,
Class. Quant. Grav. \textbf{36}, no.7, 07LT01 (2019)
[arXiv:1806.10126 [gr-qc]].

\bibitem{Pinol:2020cdp}
L.~Pinol, S.~Renaux-Petel and Y.~Tada,
[arXiv:2008.07497 [astro-ph.CO]].

\bibitem{Linde:1985gh} 
A.~D.~Linde,
Phys.\ Lett.\  {\bf 160B}, 243 (1985).

\bibitem{Affleck:1984fy} 
I.~Affleck and M.~Dine,
Nucl.\ Phys.\ B {\bf 249}, 361 (1985).

\bibitem{Dolgov:1991fr} 
A.~D.~Dolgov,
Phys.\ Rept.\  {\bf 222}, 309 (1992).

\bibitem{Dine:2003ax}
M.~Dine and A.~Kusenko,
Rev. Mod. Phys. \textbf{76}, 1 (2003)
[arXiv:hep-ph/0303065 [hep-ph]].

\bibitem{Kusenko:2014lra} 
A.~Kusenko, L.~Pearce and L.~Yang,
Phys.\ Rev.\ Lett.\  {\bf 114}, no. 6, 061302 (2015)
[arXiv:1410.0722 [hep-ph]].

\bibitem{Kusenko:2014uta}
A.~Kusenko, K.~Schmitz and T.~T.~Yanagida,
Phys. Rev. Lett. \textbf{115}, no.1, 011302 (2015)
[arXiv:1412.2043 [hep-ph]].

\bibitem{Yang:2015ida} 
L.~Yang, L.~Pearce and A.~Kusenko,
Phys.\ Rev.\ D {\bf 92}, no. 4, 043506 (2015)
[arXiv:1505.07912 [hep-ph]].

\bibitem{Pearce:2015nga} 
L.~Pearce, L.~Yang, A.~Kusenko and M.~Peloso,
Phys.\ Rev.\ D {\bf 92}, no. 2, 023509 (2015)
[arXiv:1505.02461 [hep-ph]].

\bibitem{Kusenko:2017kdr} 
M.~Kawasaki, L.~Pearce, L.~Yang and A.~Kusenko,
Phys.\ Rev.\ D {\bf 95}, no. 10, 103006 (2017)
[arXiv:1701.02175 [hep-ph]].

\bibitem{Inomata:2018htm} 
K.~Inomata, M.~Kawasaki, A.~Kusenko and L.~Yang,
JCAP {\bf 1812}, no. 12, 003 (2018)
[arXiv:1806.00123 [astro-ph.CO]].

\bibitem{Wu:2019ohx}
Y.~P.~Wu, L.~Yang and A.~Kusenko,
JHEP \textbf{12}, 088 (2019)
[arXiv:1905.10537 [hep-ph]].

\bibitem{Hook:2015foa} 
A.~Hook,
JHEP {\bf 1511}, 143 (2015)
[arXiv:1508.05094 [hep-ph]].


\bibitem{Lee:1973iz}
T.~D.~Lee,
Phys. Rev. D \textbf{8}, 1226-1239 (1973).


\bibitem{Adshead:2020ijf} 
P.~Adshead, L.~Pearce, J.~Shelton and Z.~J.~Weiner,
arXiv:2002.07201 [hep-ph].

\bibitem{Petraki:2013wwa}
K.~Petraki and R.~R.~Volkas,
Int. J. Mod. Phys. A \textbf{28}, 1330028 (2013)
[arXiv:1305.4939 [hep-ph]].

\bibitem{Hertzberg:2013mba}
M.~P.~Hertzberg and J.~Karouby,
Phys. Rev. D \textbf{89}, no.6, 063523 (2014)
[arXiv:1309.0010 [hep-ph]].

\bibitem{Takeda:2014eoa}
N.~Takeda,
Phys. Lett. B \textbf{746}, 368-371 (2015)
[arXiv:1405.1959 [astro-ph.CO]].

\bibitem{Bamba:2016vjs}
K.~Bamba, N.~D.~Barrie, A.~Sugamoto, T.~Takeuchi and K.~Yamashita,
Mod. Phys. Lett. A \textbf{33}, no.17, 1850097 (2018)
[arXiv:1610.03268 [hep-ph]].


\bibitem{Cline:2019fxx}
J.~M.~Cline, M.~Puel and T.~Toma,
Phys. Rev. D \textbf{101}, no.4, 043014 (2020)
[arXiv:1909.12300 [hep-ph]].

\bibitem{Cline:2020mdt}
J.~M.~Cline, M.~Puel and T.~Toma,
JHEP \textbf{05}, 039 (2020)
[arXiv:2001.11505 [hep-ph]].

\bibitem{Lin:2020lmr}
C.~M.~Lin and K.~Kohri,
Phys. Rev. D \textbf{102}, no.4, 043511 (2020)
[arXiv:2003.13963 [hep-ph]].

\bibitem{Babichev:2018sia}
E.~Babichev, D.~Gorbunov and S.~Ramazanov,
Phys. Lett. B \textbf{792}, 228-232 (2019)
[arXiv:1809.08108 [astro-ph.CO]].

\bibitem{Lloyd-Stubbs:2020sed}
A.~Lloyd-Stubbs and J.~McDonald,
[arXiv:2008.04339 [hep-ph]].

\bibitem{Bell:2011tn}
N.~F.~Bell, K.~Petraki, I.~M.~Shoemaker and R.~R.~Volkas,
Phys. Rev. D \textbf{84}, 123505 (2011)
[arXiv:1105.3730 [hep-ph]].

\bibitem{Dine:1995kz}
M.~Dine, L.~Randall and S.~D.~Thomas,
Nucl. Phys. B \textbf{458}, 291-326 (1996)
[arXiv:hep-ph/9507453 [hep-ph]].

\bibitem{vonHarling:2012yn}
B.~von Harling, K.~Petraki and R.~R.~Volkas,
JCAP \textbf{05}, 021 (2012)
[arXiv:1201.2200 [hep-ph]].

\bibitem{Bamba:2018bwl}
K.~Bamba, N.~D.~Barrie, A.~Sugamoto, T.~Takeuchi and K.~Yamashita,
Phys. Lett. B \textbf{785}, 184-190 (2018)
[arXiv:1805.04826 [hep-ph]].

\bibitem{Barrie:2020hiu}
N.~D.~Barrie, A.~Sugamoto, T.~Takeuchi and K.~Yamashita,
JHEP \textbf{08}, 072 (2020)
[arXiv:2001.07032 [hep-ph]].

\bibitem{Risken:1984}
H. Risken, ``The Fokker-Planck Equation'' (Springer, 1984).

\bibitem{Cohen:1987vi} 
A.~G.~Cohen and D.~B.~Kaplan,
Phys.\ Lett.\ B {\bf 199}, 251 (1987).

\bibitem{Kusenko:1997si}
A.~Kusenko and M.~E.~Shaposhnikov,
Phys. Lett. B \textbf{418}, 46-54 (1998)
[arXiv:hep-ph/9709492 [hep-ph]].





\bibitem{Kunimitsu:2012xx}
T.~Kunimitsu and J.~Yokoyama,
Phys. Rev. D \textbf{86}, 083541 (2012)
[arXiv:1208.2316 [hep-ph]].





\end{thebibliography}
\end{document}